\documentclass{appolb}
\usepackage{epsfig}

\begin{document}
\title{Theory of Soft Electromagnetic Emission in Heavy-Ion Collisions%
\thanks{Presented at 51. Cracow School of Theoretical Physics on 
``Soft Side of the LHC"}%
}
\author{Ralf Rapp
\address{Cyclotron Institute and Department of Physics \& Astronomy, 
Texas A\&M University, College Station, TX 77843-3366}
}
\maketitle
\begin{abstract}
A status report of utilizing soft electromagnetic radiation (aka thermal 
photons and dileptons) in the diagnosis of strongly interacting matter 
in ultrarelativistic heavy-ion collisions is given. After briefly 
elaborating on relations of the electromagnetic spectral function to 
chiral symmetry restoration and the transition from hadron to quark 
degrees of freedom, various calculations of electromagnetic emission 
rates in the hadronic and quark-gluon plasma phases of QCD matter are
discussed. This, in particular, includes insights from recent thermal 
lattice QCD computations. Applications to dilepton and photon spectra in 
heavy-ion collisions highlight their role as a spectro-, thermo-, baro-
and chrono-meter of extraordinary precision. 
\end{abstract}

\section{Introduction}
Collisions of heavy nuclei at (ultra-) relativistic energies provide the
fascinating opportunity to recreate blobs of strongly interacting matter
which last existed naturally almost 14 billion years ago, for a few
microseconds after the Big Bang. They are furthermore the only means by
which the bulk properties of a nonabelian gauge theory can be compared
to experiment. This is particular important in a situation where the
coupling strength is not small, implying a wealth of nonperturbative
and collective phenomena which can only be unraveled in close
collaboration between experiment and theory. 

\begin{figure}[!t]
\begin{minipage}{0.5\linewidth}
\epsfig{file=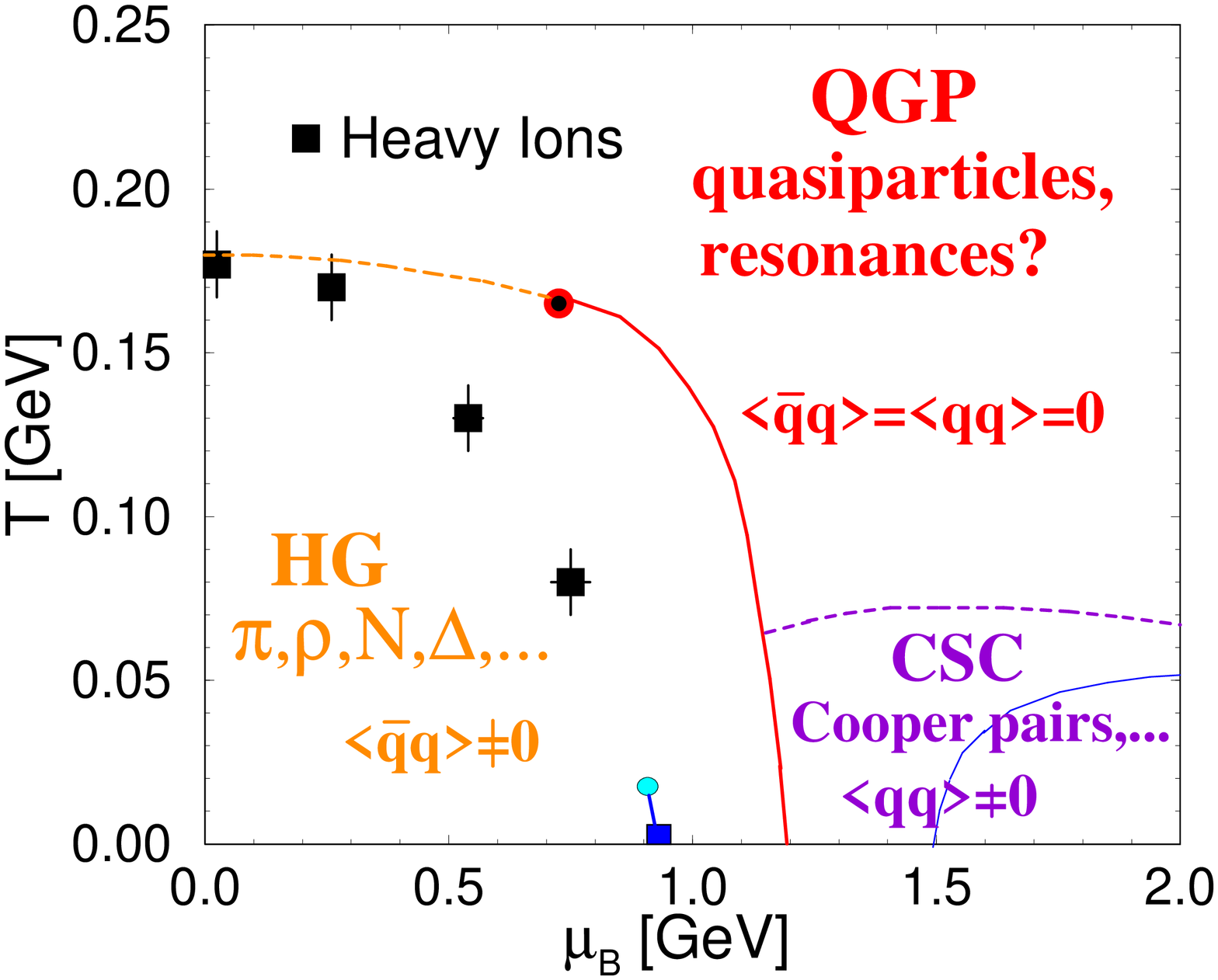,width=0.88\linewidth,angle=-90}
\end{minipage}
\hspace{0.5cm}
\begin{minipage}{0.5\linewidth}
\epsfig{file=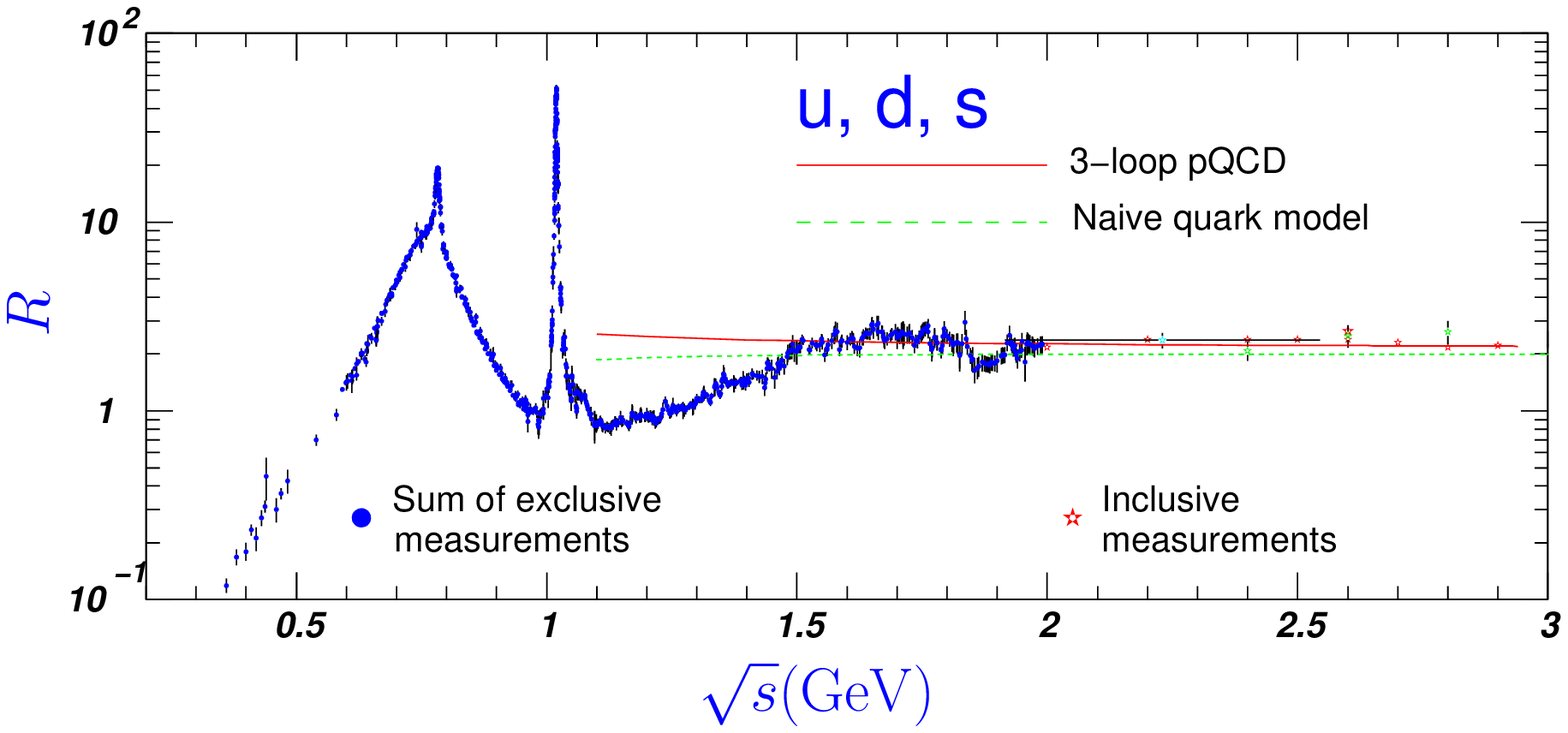,width=0.93\linewidth}
\end{minipage}
\caption{Left panel: schematic phase diagram of strong-interaction
matter, indicating a cross-over transition (dashed line) at large 
temperature, $T$, and small baryon chemical potential, $\mu_B$, and a 
first order transition into a color-superconductor (CSC) at small $T$ 
and large $\mu_B$ (solid line). The ``data points" represent ($\mu_B,T$) 
values extracted from ratios of different hadron species employing a 
thermal hadron-resonance-gas model~\cite{BraunMunzinger:2003zd}.  
Right panel: EM spectral function in vacuum as 
measured by the $R$-ratio in $e^+e^-$ annihilation into hadrons (figure 
taken from Ref.~\cite{Nakamura:2010zzi}). The EM spectral function is 
the only one whose medium modifications are directly accessible in 
heavy-ion collisions, through dilepton invariant-mass spectra.}
\label{fig_phase-dia}
\end{figure}
Over the last twenty years or so, ultrarelativistic heavy-ion collisions
(URHICs) at various laboratories around the world have demonstrated that 
systematic studies over a large region of the QCD phase diagram are 
possible, cf.~left panel of Fig.~\ref{fig_phase-dia}. For example, the 
measured hadron abundances can be well explained by a chemically 
equilibrated hadron-resonance gas including strange particles (contrary 
to $p$-$p$ collisions)~\cite{BraunMunzinger:2003zd}, and the 
transverse-momentum ($p_t$) spectra of hadrons can be well described by 
an explosive blast-wave source with common temperature and collective 
``flow" velocity in excess of half the speed of light. The observed 
azimuthal asymmetries in the $p_t$ spectra in non-central collisions 
(the so-called elliptic flow, $v_2$) can be well accounted for through 
the pressure gradients as obtained from a locally equilibrated medium 
driven by hydrodynamic expansion~\cite{Hirano:2011}. 
However, hadronic observables 
ultimately emanate from the ``freezeout" configuration of the fireball 
(typically at a temperature of $T_{\rm fo}\simeq 100$\,MeV), and thus 
do not provide information on the microscopic structure and interactions 
of medium.

Electromagnetic (EM) observables, photons and dileptons, are special
(see Refs.~\cite{Rapp:2009yu,Gale:2009gc} for recent reviews). Their 
interaction rate in the strongly interacting medium is small enough 
for them to escape the interior of the fireball unaffected, but large 
enough to be produced in measurable quantities. The radiation of photons 
from a thermalized fireball has been long recognized as powerful 
thermometer. The emission of virtual photons (aka dileptons, $e^+e^-$ 
or $\mu^+\mu^-$) carries additional information 
in terms of their invariant mass, $M=(q_0^2-\vec q^2)^{1/2}$. In fact, 
dilepton invariant-mass spectra are the only observable which gives 
direct access to the in-medium modification of a hadronic spectral 
function (strong decays like $\Delta\to\pi N$ or $\rho\to\pi\pi$ suffer 
from final state absorption and are largely emanating from the freezeout
process of the collision). This is evident from the 8-differential 
thermal production rate, 
\begin{equation}
\frac{dN_{ee}}{d^4xd^4q} = -\frac{\alpha_{\rm EM}^2}{\pi^3M^2}
\ f^B(q_0;T) \ {\rm Im}\Pi_{\rm EM}(M,q;\mu_B,T) \ ,  
\label{rate}
\end{equation}       
which only depends on the in-medium spectral function, 
Im\,$\Pi_{\rm EM}$, its thermal weight in terms of the Bose-Einstein 
distribution, $f^B$, and a free virtual-photon propagator, $1/M^2$. A 
clean temperature measurement becomes possible if Im\,$\Pi_{\rm EM}$ is 
reliably known, which is usually the case for large $M^2$ where
temperature corrections can be assessed perturbatively (in principle 
this also applies for photons at large $q_0^2=q^2$, but the leading term 
is already nontrivial in the strong coupling, i.e., 
${\cal O}(\alpha_s T^2)$, while for dileptons it is  ${\cal O}(1)$). 
However, the physics accessible through Im\,$\Pi_{\rm EM}$ is much 
richer. In the vacuum, the latter is accurately known from the inverse 
process of $e^+e^-$ annihilation into hadrons, as displayed in the 
right panel of Fig.~\ref{fig_phase-dia} in terms of the famous ratio
$R=\sigma(e^+e^-$$\to$hadrons$)/\sigma(e^+e^-$$\to$$\mu^+\mu^-)\propto
{\rm Im}\Pi_{\rm EM}^{\rm vac}/M^2$.
Its medium modifications encode a wide variety of properties of the
strongly interacting medium. For example, transport coefficients can
be extracted by the spacelike and timelike limits of vanishing energy
and momentum, corresponding to the EM susceptibility and conductivity,
respectively, of the medium. A more ambitious goal is to systematically
map out how the hadronic degrees of freedom, represented by the $\rho$, 
$\omega$ and $\phi$ resonance peaks, are affected across the QCD phase 
diagram and in this way reflect its structure. In particular,
one would like to find out how the nonperturbatively generated mass of 
the vector mesons dissolves and how they ultimately yield to a spectral
function of weakly correlated quarks at high temperatures, signaling 
deconfinement. Note that light vector mesons (especially the $\rho$)
may well serve as a prototype hadronic resonance (it is neither a 
Goldstone boson nor a heavy-quark state) in the sense that their fate
in hot/dense matter is shared by large set of resonances. A 
hadron-resonance gas is known to give a good approximation of the 
the QCD partition function until close to the phase 
transition~\cite{Karsch:2003vd}. 

\begin{figure}[!t]
\begin{minipage}{0.5\linewidth}
\epsfig{file=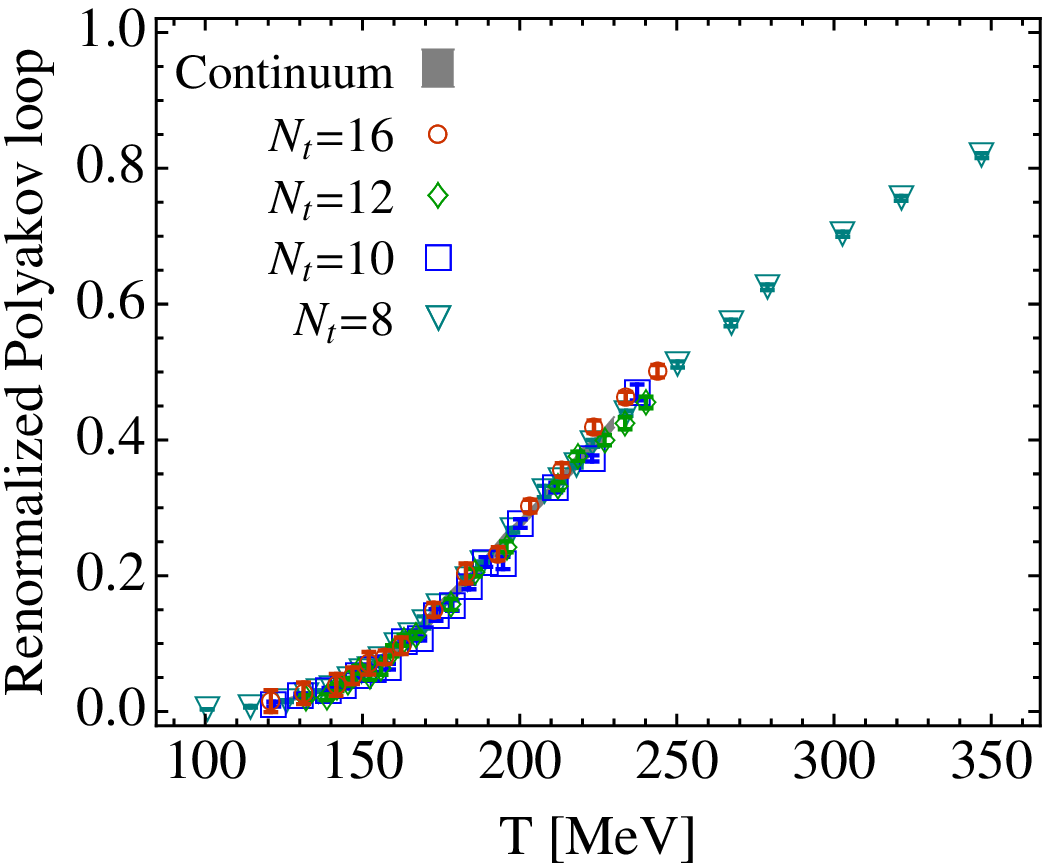,width=0.97\linewidth}
\end{minipage}
\begin{minipage}{0.5\linewidth}
\vspace{-0.1cm}
\epsfig{file=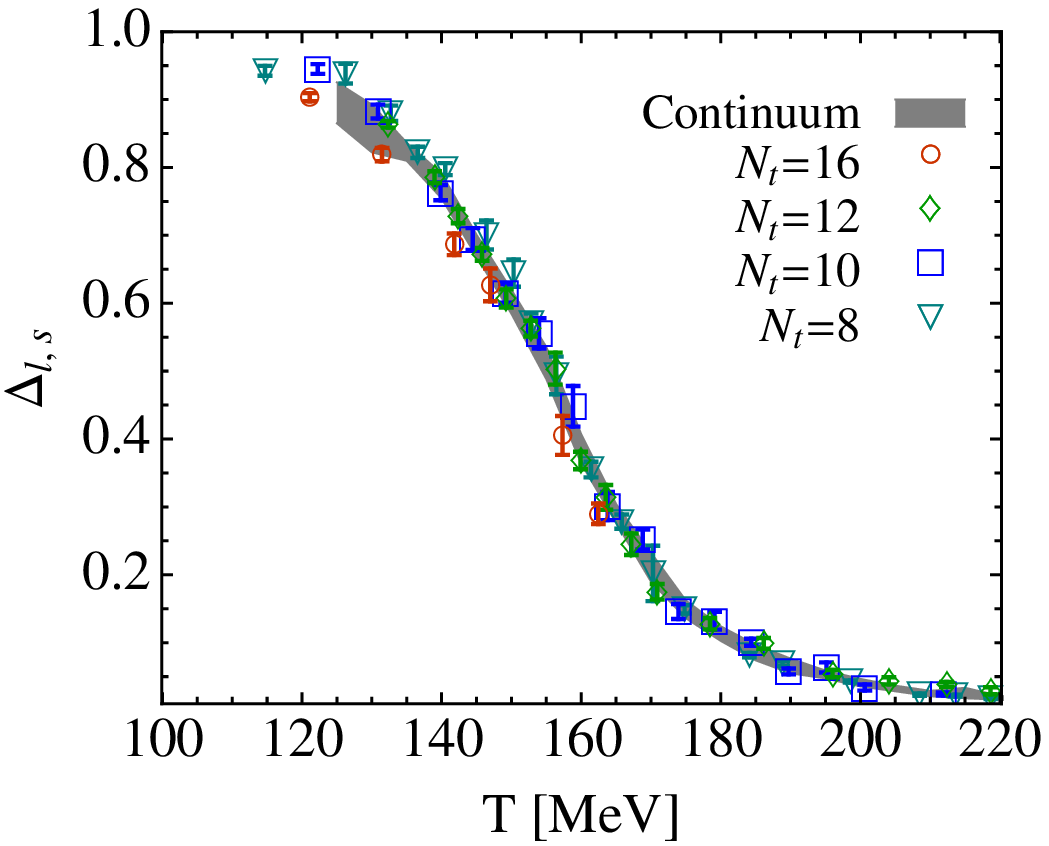,width=1.0\linewidth,angle=-0}
\end{minipage}
\vspace{0.1cm}
\caption{Results of lattice-QCD computations~\cite{Borsanyi:2010bp} for 
the renormalized Polyakov loop (left) and for the ``strangeness-subtracted" 
chiral condensate normalized to its vacuum value, 
$\langle \bar qq\rangle(T)/\langle \bar qq\rangle(0)$ (right). Both 
quantities are order parameters of the QCD phase transition, the former 
associated with quark deconfinement in the limit of large quark masses 
and the latter with chiral symmetry restoration for $m_q\to0$. The 
computations are carried out with realistic quark masses for $N_f=2+1$ 
flavors with the grey band indicating the extrapolated continuum limit.
Figures are taken from Ref.~\cite{Borsanyi:2010bp,FR-priv}.}
\label{fig_lat}
\end{figure}
Mass de-generation and deconfinement are closely related to the fate of 
the QCD condensates in the medium. Figure~\ref{fig_lat} shows recent 
results of thermal lattice QCD for order parameters of the corresponding 
QCD phase transition(s). These calculations are carried out for 
$N_f$=2+1-flavor QCD with realistic light quark masses and clearly 
demonstrate the cross-over of the transition from a chirally broken
hadronic phase to chirally restored partonic matter. It is rather 
intriguing, though, that the inflection points of these quantities,
usually associated with the pseudo-critical temperature, seem to be 
separated by about 20\,MeV, i.e., $T_c^\chi\simeq 150$\,MeV and  
$T_c^{\rm conf}\simeq170$\,MeV for chiral restoration and deconfinement,
respectively. Even without the notion of a pseudo-critical temperature,
which is not unambiguous, it is still remarkable that at, say,
$T$$\simeq$160~\,MeV the Polyakov loop has risen by merely $\sim$10\% 
toward its asymptotic value while the chiral condensate has already
dropped to below 50\%. This is suggestive for a medium in which chiral 
symmetry is largely restored but hadrons are still prevalent degrees of 
freedom (not unlike the ``quarkyonic phase" conjectured to exist in the 
moderate-$\mu_B$ region of the QCD phase diagram~\cite{McLerran:2007qj}). 
This would furthermore imply that hadronic degrees of freedom are the 
adequate basis to address mechanisms of chiral restoration. Partial 
chiral restoration in the hadronic phase is very encouraging from an 
experimental point of view, as it turns out that thermal dilepton 
emission in the low-mass region (LMR, $M\le 1$\,GeV) is dominated by 
the contribution from hot and dense matter at temperatures 
$T\simeq 150-200$\,MeV (this follows from the interplay of the 
increasing 3-volume and the decreasing thermal Bose factor in the 
space-time integrated spectral yield, see Sec.~\ref{ssec_fb} 
below)~\cite{Rapp:2004zh}.

\begin{figure}[!t]
\begin{center}
\epsfig{file=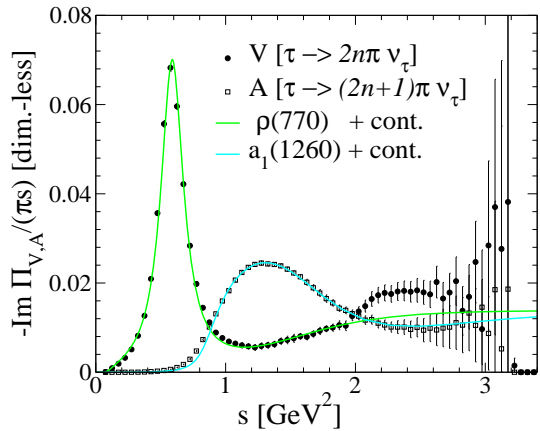,width=0.55\linewidth,angle=-0}
\end{center}
\caption{Vacuum spectral functions of the vector and axialvector 
currents as measured in hadronic $\tau$ decays into Goldstone 
bosons~\cite{Barate:1998uf,Ackerstaff:1998yj}. The lines are model
calculations using $\rho$ and $a_1$ dominance with smoothly onsetting
multi-pion continua.}
\label{fig_VA-vac}
\end{figure}
The manifestation of the Spontaneous Breaking of Chiral Symmetry (SBCS) 
in the vacuum EM spectral function becomes apparent upon decomposing
it into good isospin states. The dominant contribution arises from
the isovector channel ($I$=1) which in the LMR is essentially saturated
by the $\rho$ meson (the contribution of the isoscalar-vector ($\omega$) 
channel to $\Pi_{\rm EM}$ is down by a factor of $\sim$10). 
The Weinberg sum rules~\cite{Weinberg:1967,Das:1967ek} relate 
moments of the isovector-vector and -axialvector spectral functions to 
order parameters of SBCS, 
\begin{equation}
f_n = - \int\limits_0^\infty \frac{ds}{\pi} \ s^n \
 \left[{\rm Im} \Pi_V(s) - {\rm Im} \Pi_A(s) \right]  \ . 
\label{wsr}
\end{equation} 
Both spectral functions have been accurately measured in vacuum via 
hadro\-nic decays of the $\tau$ lepton at 
LEP~\cite{Barate:1998uf,Ackerstaff:1998yj} (see Fig.~\ref{fig_VA-vac})
and constitute one of the best experimental
evidences for SBCS. For example, for $n=-1$, one has $f_{-1}=f_\pi^2$,
where $f_\pi$=93\,MeV is the pion decay constant (or pion 
``polestrength"). Recalling the Gellmann-Oakes-Renner relation,
$f_\pi^2 m_\pi^2 = - m_q  \langle0|\bar qq|0\rangle$, one recognizes
the close relation between scalar quark condensate, Goldstone bosons
and the isovector axial-/vector spectral functions (the explicit chiral
breaking is signified by the small current quark mass, 
$m_q\simeq5$\,MeV, and the pion mass). The Weinberg sum rules, which 
remain valid in the medium, exhibit the important role of the axialvector 
spectral function in the search for chiral restoration from dilepton 
data: the vector spectral function degenerates with its chiral
partner, the axialvector, in the chirally restored phase.
Similar to the vector channel, the nonperturbative part of the 
axialvector channel is dominated by a resonance, the $a_1(1260)$.
At higher masses, $M>1.5$\,GeV, one expects from perturbation theory 
that the axialvector also merges into a continuum which coincides
with the vector channel.
Another example of relating QCD condensates to empirical spectral
are QCD sum rules whose application to the $\rho$ meson will be
discussed below.

The remainder of this article is organized as follows. 
In Sec.~\ref{sec_vc} we introduce the vector-current correlator in 
strongly interacting matter and recall its non-interacting and vacuum
limits (Sec.~\ref{ssec_vac}), review calculations of its spectral 
function in hadronic (Sec.~\ref{ssec_had}) and partonic media including 
recent results from thermal lattice QCD (Sec.~\ref{ssec_qgp}), 
and discuss the corresponding thermal production rates of dileptons 
and photons (Sec.~\ref{ssec_rates}). In Sec.~\ref{sec_hics} we turn to 
the quantitative analysis of EM emission spectra in URHICs, beginning
with descriptions of the space-time evolution of the medium over
which the rates need to be integrated (Sec.~\ref{ssec_fb}).  
Theoretical calculations of EM spectra in URHICs will be compared
to experiment to extract information on medium effects of the spectral 
function, the time duration of emission (spectro- and chrono-meter, 
Sec.~\ref{ssec_spec}) and temperature and collective properties of the 
expanding fireball (thermo- and baro-meter, Sec.~\ref{ssec_thermo}).
We conclude in Sec.~\ref{sec_concl}.

\section{Vector-Current Correlator in Medium}
\label{sec_vc}
In this section the central role is played by the hadronic correlation 
function of two electromagnetic currents defined as
\begin{equation}
\Pi_{\rm EM}^{\mu\nu}(q) = -i  \int d^4x \ {\rm e}^{iqx} \ \Theta(x_0) 
\ \langle [j^\mu_{\rm EM}(x), j^\nu_{\rm EM}(0)]\rangle_T \ , 
\label{Pi-em}
\end{equation} 
where $\langle\cdots\rangle_T$ denotes the expectation value at finite
temperature (giving rise to a retarded spectral function).
In a partonic basis, corresponding to the elementary degrees of 
freedom in the QCD Lagrangian, the EM current takes the form
\begin{equation}
j^\mu_{\rm EM} = \frac{2}{3} \bar u \gamma^\mu u - 
  \frac{1}{3} \bar d \gamma^\mu d -\frac{1}{3} \bar s \gamma^\mu s
\label{j_quark}
\end{equation} 
where we constrain ourselves to the 3 light flavors up, down and strange. 
The EM current in quark basis, Eq.~(\ref{j_quark}), can be rearranged 
into good isospin states which naturally leads to the hadronic basis 
according to
\begin{eqnarray}
j^\mu_{\rm EM} &=& \frac{1}{2} (\bar u \gamma^\mu u -\bar d \gamma^\mu d)
                  +\frac{1}{6} (\bar u \gamma^\mu u +\bar d \gamma^\mu d)
                  -\frac{1}{3}  \bar s \gamma^\mu s
\nonumber\\
  &=& \frac{1}{\sqrt{2}} j_\rho^\mu  + \frac{1}{3\sqrt{2}} j_\omega^\mu
     -\frac{1}{3} j_\phi^\mu
\label{j_had}
\end{eqnarray}
with the properly normalized hadronic currents, $j_v^\mu$ 
($v=\rho, \omega, \phi$). Invoking the vector-dominance model (VDM)
characterized by the field-current identity $j_v^\mu=m_v^2/g_v$, the 
hadronic currents are saturated by the light vector mesons and the 
pertinent current-current correlator turns into a vector-meson 
propagator, $D_v$.

\subsection{EM Spectral Function in Vacuum and Thermal Emission Rates}
\label{ssec_vac}
The imaginary part of the EM current-current correlator is the EM
spectral function. In the vacuum it is directly observable in 
$e^+e^-$ annihilation, recall right panel of Fig.~\ref{fig_phase-dia}.
Its structure suggests a decomposition into a nonperturbative low-mass
regime, $M\le 1$\,GeV, saturated by hadronic resonances, a ``dip" region
for 1\,GeV$<M\le 1.5$\,GeV, and regime for $M>1.5$\,GeV
where perturbation theory applies, 
\begin{equation}
{\rm Im}\Pi_{\rm EM}^{\rm vac}(M) = \left\{
\begin{array}{ll}
 \sum\limits_{v=\rho,\omega,\phi} \left(\frac{m_v^2}{g_v}\right)^2 \
{\rm Im} D_v^{\rm vac}(M) & , \ M < M_{\rm dual}^{\rm vac} ,
\vspace{0.3cm}
\\
-\frac{M^2}{12\pi} \ (1+\frac{\alpha_s(M)}{\pi} +\dots)  \ N_c
\sum\limits_{q=u,d,s} (e_q)^2  & , \ M > M_{\rm dual}^{\rm vac} \
\end{array}
\right.
\label{ImPi-em}
\end{equation}
($N_c$=3 is the number of colors and $e_q$ the quark charge in units of 
the electron charge).
Note that in the perturbative regime the final state in $e^+e^-$ 
annihilation is still hadronic (with broad overlapping excited resonances 
such as $\rho'$, $\omega'$, etc.), but the ``long-distance" hadronization 
process does not affect the short-distance quark-production cross section. 
This is often referred to as parton-hadron duality, which in the vector 
channel is realized for masses beyond a ``duality threshold", 
$M>M_{\rm dual}^{\rm vac}\simeq1.5$\,GeV.

The inverse process of $e^-e^+$ annihilation is intimately related to
the production rate of thermal dileptons from a heat bath of strongly 
interacting matter. It can be calculated starting from Fermi's Golden
rule, 
\begin{eqnarray}
\frac{\Gamma_{ee}}{V}=\frac{dN_{ee}}{d^4x} &=& e^4 \int 
\prod_i\frac{d^3p_i}{(2\pi)^3 2E_i} \prod_f\frac{d^3p_f}{(2\pi)^3 2E_f}
2\pi \delta(P_i-P_f-q) 
\nonumber\\
&&\times\langle i|j^{\rm em}_\mu|f\rangle \langle f|j^{\rm em}_\nu|i\rangle 
f^i (1\pm f^f) \left(j_e^\mu\frac{1}{q^4} j_e^\nu \right) \ , 
\end{eqnarray} 
which can be cast into the form given in Eq.~(\ref{rate}). 
This expression is leading order in the fine structure constant,
${\cal O}(\alpha_{\rm EM}^2)$, but exact in the strong interactions 
which are encoded in the spectral function (the leading order in the 
strong interaction is ${\cal O}(1)$). In the same framework one can 
calculate the thermal production rate of real photons as
\begin{equation}
q_0\frac{dN_{\gamma}}{d^4xd^3q} = -\frac{\alpha_{\rm EM}}{\pi^2}
\ f^B(q_0;T) \ {\rm Im}\Pi_{\rm EM}^T(M=0,q=q_0;\mu_B,T) \ ,
\label{Rgam}
\end{equation}
which only involves the transverse polarization of the EM spectral function.
The leading orders of the photon rate are one lower in the EM coupling, 
${\cal O}(\alpha_{\rm EM})$, but one higher in the strong coupling,
${\cal O}(\alpha_s)$, relative to the dilepton rate. This implies 
significant uncertainties in the magnitude of the rate even in the
perturbative regime. 

In the following three sections we focus on the dilepton rate in the
soft (low-mass) regime, where the key issue is to determine the 
in-medium EM spectral function.

\subsection{Hadronic Matter}
\label{ssec_had}
At low temperatures and densities, the appropriate degrees of freedom
to describe strongly interacting matter are hadrons. In lukewarm matter,
$T<m_\pi$, at vanishing baryon chemical potential, $\mu_B$, the most 
abundant species are pions whose low-energy interactions are determined  
by the chiral Lagrangian. At higher temperatures, $T\ge m_\pi$, 
hadronic resonances quickly outnumber pions; much less is known about 
the chiral structure of resonance interactions (including excited nucleon 
states). However, it is still possible to write effective hadronic
Lagrangians compatible with basic symmetry requirements and constrain
their parameters by empirical information, e.g., hadronic and radiative
decay widths. The such constrained interactions can then be implemented
into many-body theory to calculate in-medium selfenergies and spectral
functions of hadronic excitations. Clearly, the reliability of the 
predicted spectral functions largely hinges on the quality of the
constraints applied to the effective interactions.   

\begin{figure}[!t]
\epsfig{file=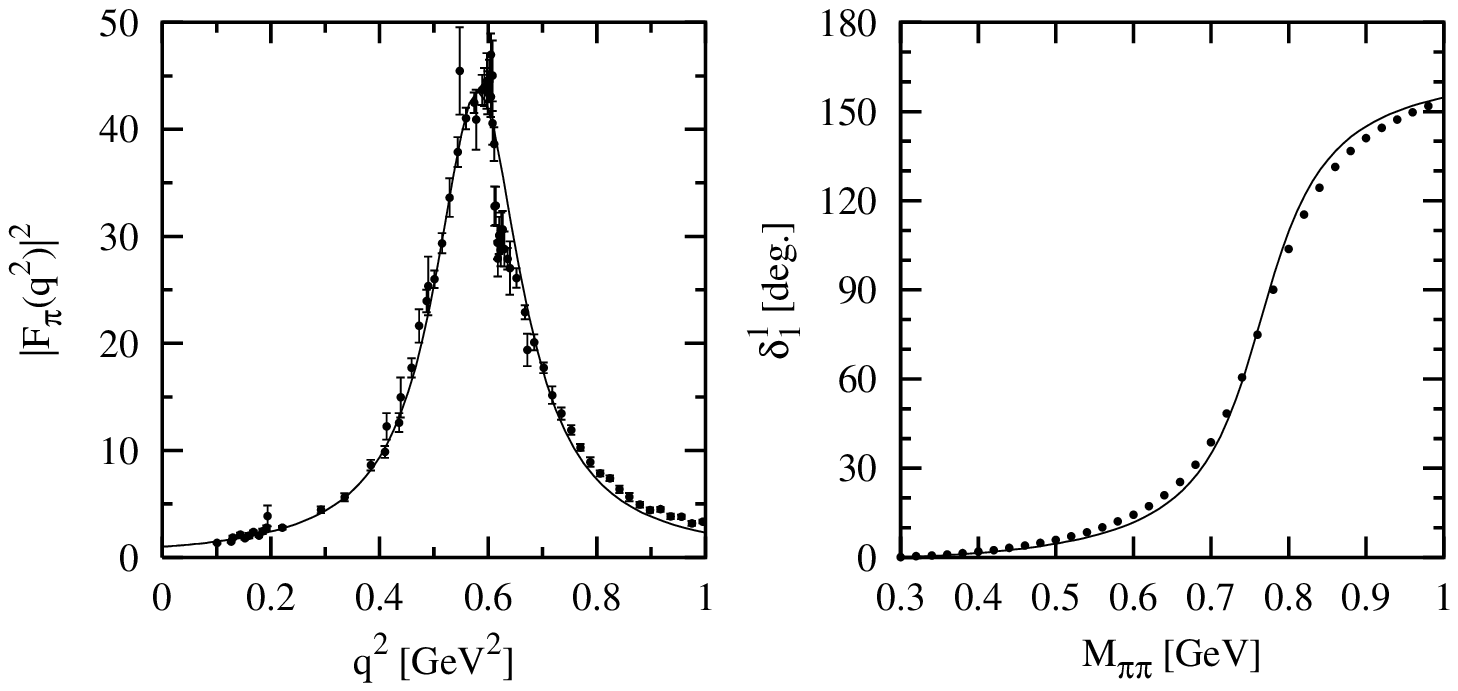,width=0.9\linewidth,angle=-0}
\caption{Electromagnetic formfactor of the pion (left panel) and $P$-wave
$\pi\pi$ scattering phase shifts (right panel) in the vacuum. An effective
model assuming $\rho$-meson dominance with coupling to two-pion states
is compared to data. Figures taken from Ref.~\cite{Urban:1998eg}.}
\label{fig_rho-vac}
\end{figure}
The above strategy has been adopted by several groups to compute the 
in-medium spectral function of the $\rho$ meson (cf., e.g., the review 
papers 
\cite{Rapp:1999ej,Alam:1999sc,Rapp:2009yu,Leupold:2009kz,Rapp:2011zz} 
for a detailed discussion and further references). 
The special role of the $\rho$ meson simply derives from its dominant
contribution to the low-mass EM correlators where it outshines the
$\omega$ by a factor of $\sim$10 (the constituent-quark model predicts
a factor of 9, as inferred from Eq.~(\ref{j_had}); from the empirical 
dilepton decay widths one finds $\Gamma_{\rho\to ee}/\Gamma_{\omega\to ee}
=(m_\rho^4/g_\rho^2)/(m_\omega^4/g_\omega^2)\simeq$11).
Starting point is a realistic description of the $\rho$ properties
in the vacuum. Usually the $\rho$ is introduced into the chiral 
Lagrangian via a local gauge symmetry. To one loop order, the
vacuum $\rho$ selfenergy is built up from the standard $\rho\to\pi\pi$ 
diagram and a tadpole arising from the $\rho\rho\pi\pi$ 4-point vertex
(the pertinent diagram has no imaginary part). With 3 free parameters
(bare $\rho$ mass, gauge coupling $g$ and a regulator (cutoff)
for the divergent loop integrals), a satisfactory reproduction of the
$\rho$ spectral shape as seen in the pion-EM formfactor and $P$-wave
$\pi\pi$ scattering can be achieved, cf.~Fig.~\ref{fig_rho-vac} (the
$\tau\to2\pi\nu_\tau$ data are also well described, see 
Fig.~\ref{fig_VA-vac}). A good description of the entire spectral 
shape -- not just its mass and width -- is essential for later 
purposes of calculating low-mass dilepton spectra, especially below 
the free $\rho$ mass. 

\begin{figure}[!t]
\vspace{-1cm}
\epsfig{file=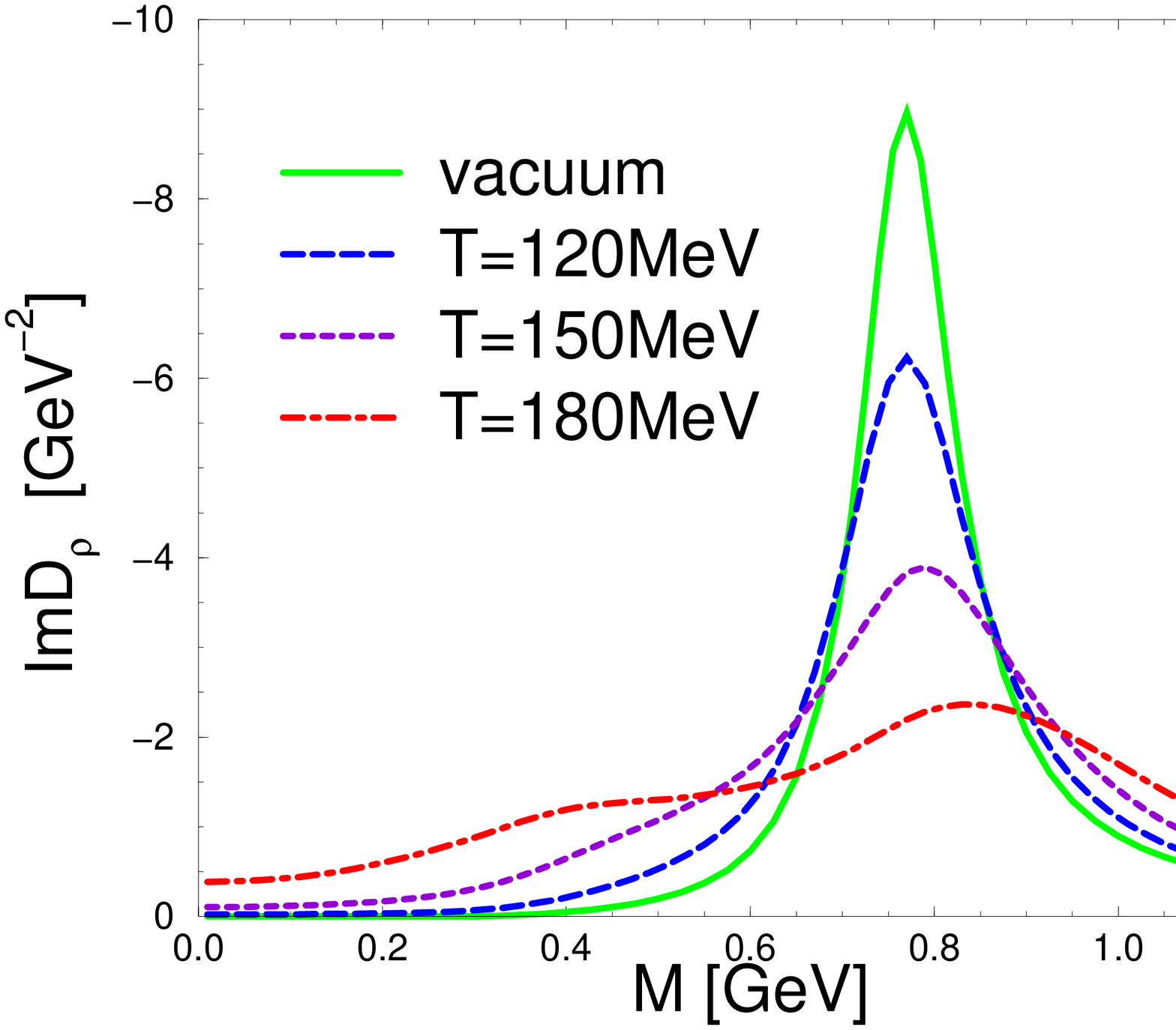,width=0.45\linewidth,angle=-0}
\epsfig{file=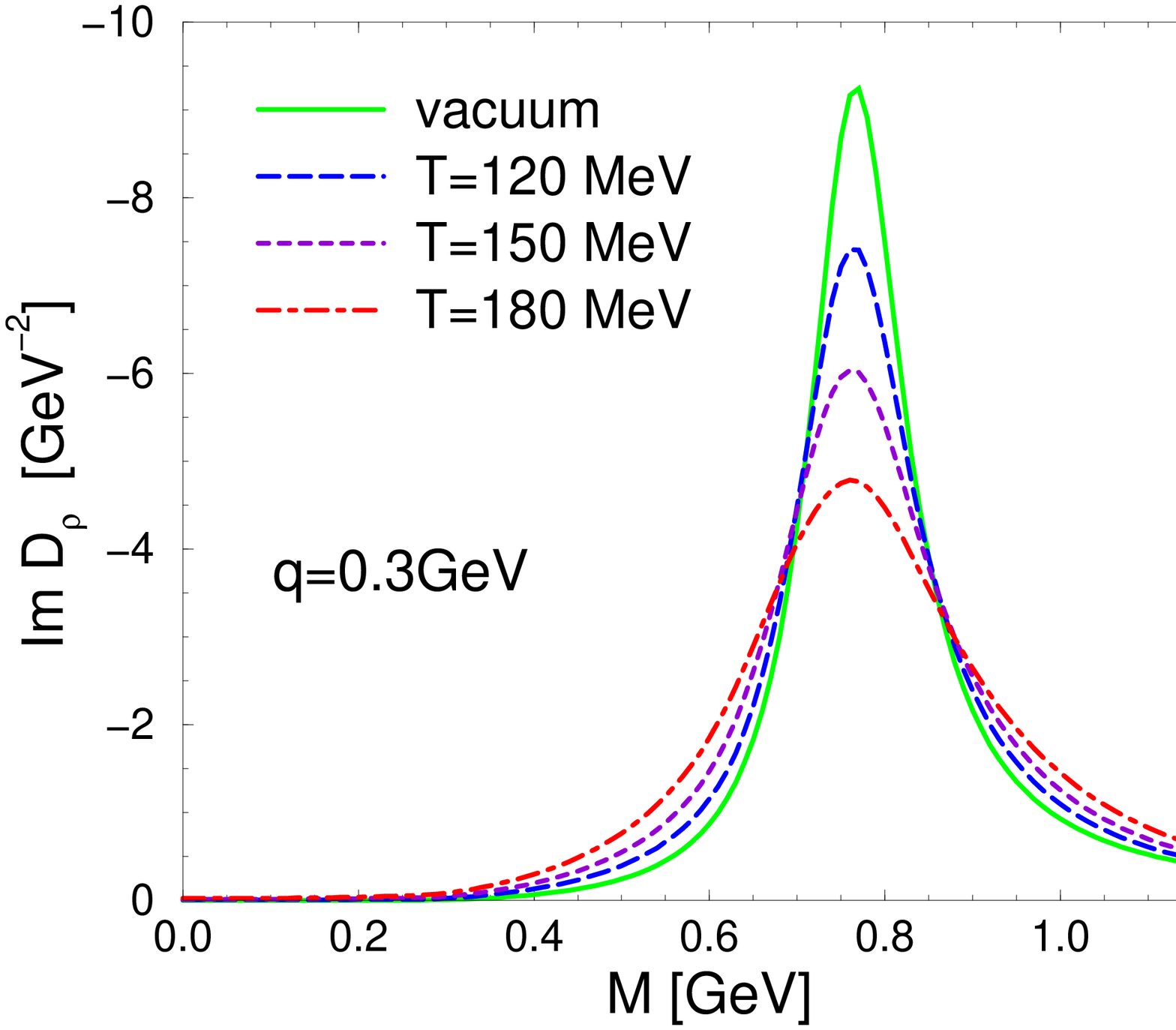,width=0.49\linewidth,angle=-0}
\caption{Spectral function of the $\rho$ meson in hot hadronic matter;
left panel: in hot and dense matter for temperatures $T$=120, 150 and
180\,MeV and at fixed baryon chemical potential of $\mu_B$=330\,MeV,
corresponding to baryon densities of 0.1$\varrho_0$, 0.7$\varrho_0$ and
2.6$\varrho_0$, respectively ($\varrho_0$=0.16\,fm$^{-3}$); right panel:
in a hot meson with all baryon-induced medium effects switched off.}
\label{fig_drho}
\end{figure}

Medium modifications of the $\rho$ propagator,
\begin{equation}
D_\rho(M,q;\mu_B,T) =
\frac{1}{M^2-m_\rho^2-\Sigma_{\rho\pi\pi}-\Sigma_{\rho M}-\Sigma_{\rho B}}
 \ ,
\label{Drho}
\end{equation}
are induced by (a) interactions of its pion cloud with hadrons from
the heat bath, included in $\Sigma_{\rho\pi\pi}$ (e.g., 
$\pi N\to \Delta$), and (b) direct $\rho$ scattering off mesons (e.g., 
$\rho\pi\to a_1$) and baryons (e.g., $\rho N\to N^*$), denoted by
$\Sigma_{\rho M}$ and $\Sigma_{\rho B}$, respectively.    
The medium effects in nuclear matter can be comprehensively constrained 
by analyzing $\pi N\to \rho N$ scattering and photoabsorption spectra
on the nucleon and nuclei~\cite{Rapp:1997ei}. In this context, the pion 
cloud effects correspond to meson-exchange currents (e.g., pion exchange 
in $\gamma N\to\gamma\Delta$), and the direct nucleon excitations, 
$\gamma N\to N^*,\Delta^*$, produce the resonance peaks in the cross 
section. For the $\rho$ interactions in a meson gas, one has to resort 
to the hadronic and radiative decay branchings of mesonic resonance 
with large coupling to $\rho$ and $\gamma$ final states, e.g., 
$a_1\to\pi\rho,\pi\gamma$ or $K_1\to K\rho,K\gamma$~\cite{Rapp:1999qu}.   
To date, for the case of cold nuclear matter, different calculations
of the in-medium $\rho$ spectral functions have reached agreement
at a semi-quantitative level, and remaining discrepancies can be
largely traced back to slight variations in the treatment of
medium effects (e.g., missing pion-cloud effects or linear-in-density
approximations).
An example for the $\rho$ spectral functions under hot and dense 
conditions as expected for heavy-ion collisions at the CERN-SPS is 
displayed in the left panel of Fig.~\ref{fig_drho}. The $\rho$ 
resonance peak undergoes a strong broadening, indicative for its 
ultimate ``melting" close to the phase transition. The medium
modifications, especially the low-mass enhancement, are much reduced 
if baryon-induced effects are switched off, see right panel of 
Fig.~\ref{fig_drho}.  

\begin{figure}[!t]
\begin{center}
\epsfig{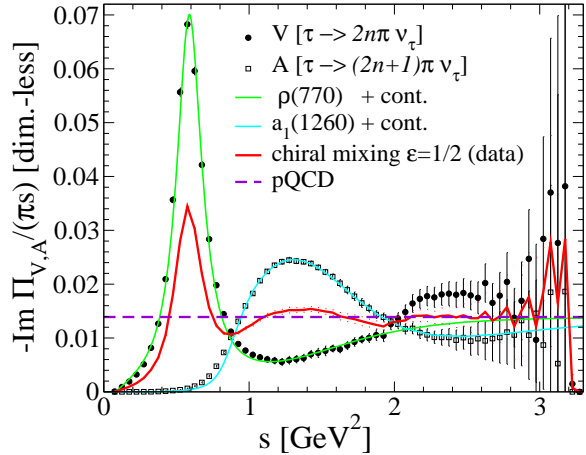}
\end{center}
\caption{Effect of chiral mixing~\cite{Dey:1990ba} on isovector-vector
($V$) and -axialvector ($A$) spectral functions. When extrapolated to 
the chiral restoration point ($\varepsilon$=1/2), the degenerate $V$
and $A$ spectral functions also degenerate with the perturbative 
continuum (dashed line) down to $s\simeq1$\,GeV$^2$, thus filling
the ``dip-region" in the vacuum $V$ spectral function.}
\label{fig_VA-mix}
\end{figure}

Let us briefly allude to medium effects in the dip region, i.e., for
masses between 1 and 1.5\,GeV. In this regime the continuum starts
to develop, characterized by multi-meson contributions in the
spectral function whose interactions with a medium are difficult
to assess microscopically. Fortunately, one can make a more simple
yet elegant argument based on chiral symmetry to estimate the
medium effects on the spectral function. It was first developed in
Ref.~\cite{Dey:1990ba} for the finite-temperature case. Using current
algebra in the chiral limit ($m_\pi=0$), the interactions of the vector 
and axialvector correlators with a lukewarm pion gas were shown to 
result in their mutual mixing as  
\begin{equation}
\Pi_{V}(q) = (1-\varepsilon)~\Pi_{V}^{\rm vac}(q) +
\varepsilon~\Pi_{A}^{\rm vac}(q) \
\label{chi-mix}
\end{equation}
(and likewise for the axialvector upon exchanging $V\leftrightarrow A$),
with the mixing parameter $\varepsilon$=$T^2/6f_\pi^2$. The
model-independent leading-temperature effect on the vector spectral
function is an interaction with a thermal pion which reduces the 
strength at its resonance and moves it into the axialvector resonance.
Such processes are of course included in the many-body treatment of the 
low-mass region discussed above, e.g., via $\rho\pi\to a_1$ interactions. 
Here we use the mixing effect to predict that the dip region in the
vector spectral function will be enhanced due to the admixture of the 
axialvector resonance ($a_1$). In fact, if one takes the $\tau$ decay 
data for the vacuum $V$ and $A$ spectral functions and carries the 
mixing all the way to the degeneracy point ($\varepsilon$=1/2), one 
finds the intriguing result that the dip region is filled ``precisely" 
to the level of the perturbative continuum, cf.~Fig.~\ref{fig_VA-mix}. 
Stated differently, chiral mixing induces the reduction of the in-medium
duality threshold to $M_{\rm dual}\simeq1$\,GeV. The mixing effect is 
not enough to smear the $\rho$ resonance -- this is where 
the (higher-order) hadronic many-body effects come in, as we will
see below.
Similar arguments can be made for nuclear matter at zero temperature,
where the role of the thermal pions is taken over by the virtual
pion cloud of the nucleon~\cite{Krippa:1997ss,Chanfray:1999me}; this
is closely related to the pion cloud effects on the $\rho$
meson through $\pi N$ scattering discussed above.

\subsection{Quark-Gluon Plasma and Lattice QCD}
\label{ssec_qgp}
In a quark-gluon plasma (QGP) the leading source of thermal dileptons is 
the purely electromagnetic annihilation of two quarks of equal flavor, 
$q\bar q\to e^+e^-$, corresponding to a structureless noninteracting EM 
spectral function, ${\rm Im}\Pi_{\rm EM}\propto M^2$, as given by the 
lower line in Eq.~(\ref{ImPi-em}). A consistent (gauge-invariant)
calculation of loop corrections requires a careful treatment of infrared
physics at the scale $gT$ ($g=\sqrt{4\pi\alpha_s}$) which can contribute 
at the same order as the leading term (due to soft propagators of the 
type $1/(t-m_D^2)$ with $m_D\sim gT$, canceling the extra coupling 
constants in the vertices). A systematic way to resolve these problems 
is provided by resummations within hard-thermal-loop (HTL) perturbation 
theory, which has been first applied to thermal dilepton production in 
Ref.~\cite{Braaten:1990wp}. The softening of the in-medium quark 
dispersion relation, developing plasmino branches as well as Landau 
damping, in the quark-antiquark loop produce an appreciable enhancement 
of the HTL dilepton rate
(or EM spectral function) over the free one, not unlike the medium
effects in the pion cloud of the $\rho$. The enhancement is present for 
a rather wide regime around $q_0\sim gT$, and growing with decreasing 
$q_0$.  

\begin{figure}[!t]
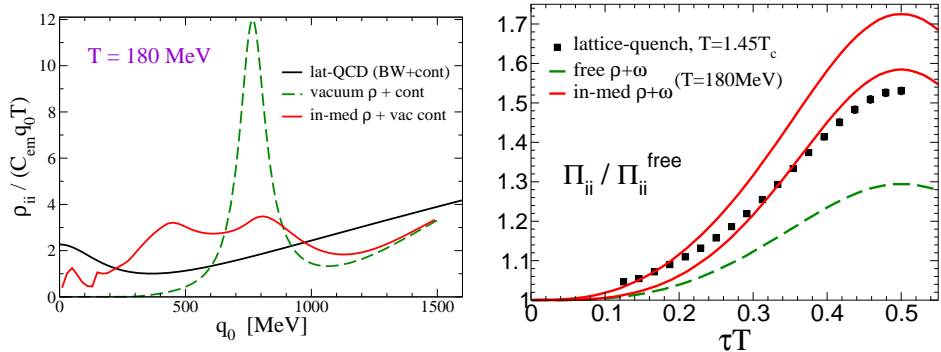

\begin{minipage}{0.5\linewidth}
\epsfig{file=rho-em-3.eps,width=0.95\linewidth}
\end{minipage}
\begin{minipage}{0.5\linewidth}
\epsfig{file=Pii180.eps,width=0.93\linewidth}
\end{minipage}
\vspace{0.1cm}
\caption{Left panel: EM spectral function, at vanishing 3-momentum and
summed over polarization, from thermal lQCD at $T$=1.45\,$T_c$ (solid
black line)~\cite{Ding:2010ga}, as well as for vacuum and in-medium 
hadronic calculations at $T$=180\,MeV (dashed and red solid line, 
respectively)~\cite{Rapp:1999us}. The spectral
functions are rendered dimensionless upon division by energy, 
temperature and a charge/isospin degeneracy ($C_{em}=\sum_{u,d}e_q^2=5/9$ 
for lQCD and $C_{em}=1/2$ for the isovector hadronic results).
Right panel: euclidean correlators corresponding to the spectral 
functions in the left panel (squares: lQCD; solid lines: in-medium 
hadronic with the free continuum (lower line) and one with a 
threshold lowered by 0.3\,GeV (upper line); dashed line: vacuum;
the hadronic calculations additionally include the contribution
from free or in-medium $\omega$ spectral functions).}
\label{fig_corr}
\end{figure}

In more recent years, the problem of soft dilepton emission has been 
addressed from first principles using lattice QCD 
(lQCD)~\cite{Ding:2010ga}, by computing 
the thermal expectation value of current-current correlation in 
Eq.~(\ref{Pi-em}) for Euclidean time, $\tau$, as
\begin{equation}
\Pi_{\mu\nu}(\tau,\vec q) = \int d^3x \ \Pi_{\mu\nu}(\tau,\vec x)
         \ {\rm e}^{i\vec p\cdot\vec x} \ .
\label{Pi-tau1}
\end{equation}
In a heat bath in equilibrium, the imaginary-time coordinate, $\tau$, is 
not related to dynamical but statistical properties of the system: the
temperature comes in through a periodicity in $\tau$,
effectively limiting it to the ``Matsubara circle", 
$0\le \tau \le 1/T$. The connection to the timelike (observable) 
spectral function can be established by means of a Kramers-Kronig 
dispersion relation leading to  
\begin{equation}
\Pi(\tau,q;T)=
\int\limits_0^\infty \frac{dq_0}{2\pi} \ \rho(q_0,q;T) \
\frac{\cosh[q_0(\tau-1/2T)]}{\sinh[q_0/2T]} \ 
\label{Pi-tau2}
\end{equation}
where $\rho(q_0,q;T)=-2~{\rm Im} \Pi^i_i$ is the polarization-summed
spectral function (in the vector channel, charge conservation implies
the 00-component to proportional to $\delta(q_0)$). 
Equation~(\ref{Pi-tau2}) illustrates a notorious problem of extracting 
spectral functions from Euclidean correlators computed in lQCD: one
needs to perform an inverse integral transform on a finite number
of meshpoints in $\tau$, which additionally carry statistical and
systematic uncertainty and are limited to the Matsubara circle. 
Nevertheless, probabilistic methods (known as maximum entropy method, 
successfully used, e.g., in image reconstruction) can be applied which 
are usually coupled with additional physical constraints such as 
positivity of the spectral function~\cite{Asakawa:2000tr}. However, it 
currently remains unclear as to how much structure can be resolved in 
the spectral functions (e.g., resonance peaks vs. threshold 
enhancements).  In Ref.~\cite{Ding:2010ga}, this problem has been 
mitigated by making a physically motivated ansatz for the EM spectral
function and fitting its 3 free parameters to the euclidean correlator 
``data". The ansatz consists of a Breit-Wigner ``transport" peak at low 
energies plus a pertubative continuum with $\alpha_s$ correction (plus
Pauli-blocking factor) at high energies, as in the lower line of 
Eq.~(\ref{ImPi-em}). The resulting spectral function in a gluon plasma 
of temperature 1.45\,$T_c$ (without thermal anti-/quarks) 
is shown as the solid black line in left panel of Fig.~\ref{fig_corr}. 
It is divided by the energy variable to exhibit the ``transport peak" 
for $q_0\to0$ which is dictated by the proper low-energy limit of the 
retarded correlator, 
$\rho_{EM}^{ii}(q_0\to0;T)\propto \sigma_{\rm EM}(T) \, q_0$,
where $\sigma_{\rm EM}$ denotes the electrical conductivity (a 
transport coefficient of the  strongly interacting medium).
Also shown are the vacuum correlator and the results of
the hadronic many-body calculations predicting a strongly broadened
$\rho$ meson at a temperature close to $T_c$. Despite the seemingly
quite different environments (gluon plasma vs. interacting hadron gas)
the two in-medium calculations are very different from the vacuum
shape, but {\em not} very different from each other.  
In fact, a cleaner comparison can be done by going back to
the euclidean correlators, by simply performing the integral over
the hadronic many-body spectral function in Eq.~(\ref{Pi-tau2}).
The resulting correlators (normalized to the non-interacting one)
for the vacuum and in-medium spectral function are compared to the
direct lQCD computations in a gluon plasma in the right panel of
Fig.~\ref{fig_corr}. The in-medium spectral function leads to a
correlator ratio quite similar to the one from lQCD, while the lack of a 
low-mass enhancement in the vacuum spectral function entails a ca.~50\% 
underestimate of the maximum in $\Pi(\tau=1/2T)$ (the euclidean 
correlators are symmetric about the midpoint of the Matsubara circle, 
$\tau=1/2T$). The rather good agreement of the in-medium correlators 
reiterates that the latter are not particularly sensitive to detailed 
structures in the spectral function, but that they do contain valuable 
information, such as a redistribution of strength to low energies
or estimates for the transport peak.

\subsection{In-Medium Diletpon and Photon Rates}
\label{ssec_rates}
\begin{figure}[!t]
\begin{minipage}{0.55\linewidth}
\epsfig{file=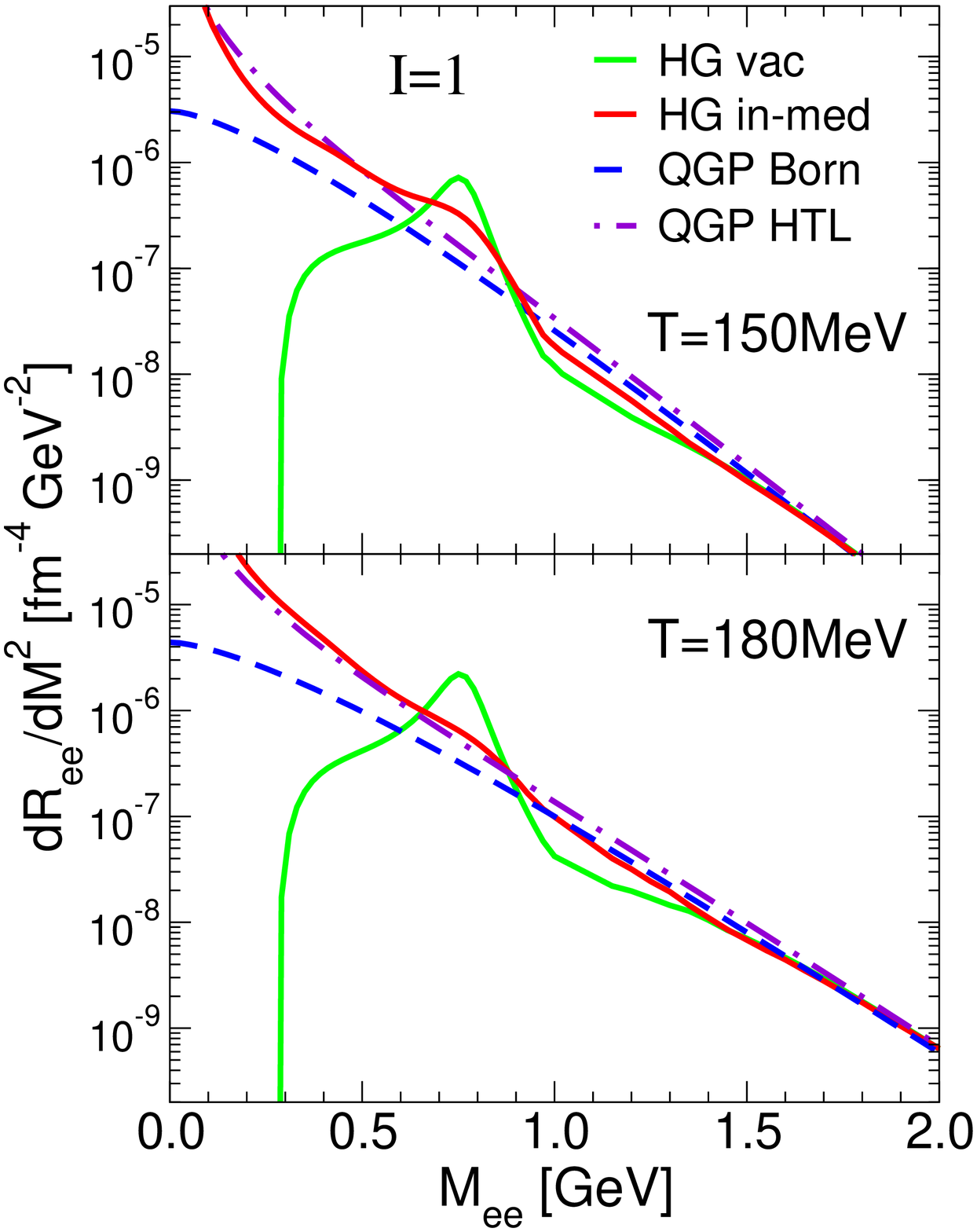,width=0.95\linewidth}
\end{minipage}
\hspace{-0.8cm}
\begin{minipage}{0.55\linewidth}
\epsfig{file=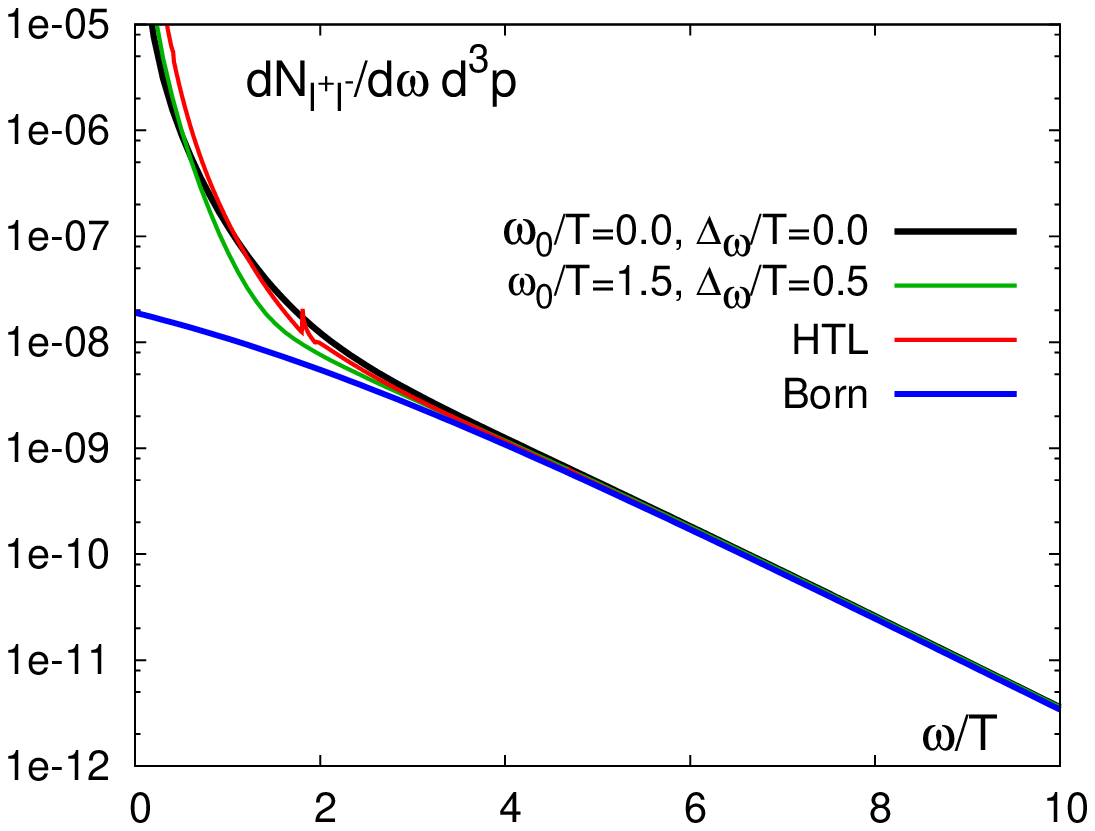,width=0.95\linewidth}
\end{minipage}
\caption{Thermal dilepton rates; left panel: from a hadron gas with 
vacuum (solid green line) and in-medium~\cite{Rapp:1999us} (solid red 
line) EM spectral function, compared to $q\bar q$ annihilation in 
leading-order (dashed line) and with hard-thermal loop 
corrections~\cite{Braaten:1990wp} (dashed-dotted line);  all rates
are integrated over the pair 3-momentum; right panel: from quenched 
lattice QCD at 1.45\,$T_c$~\cite{Ding:2010ga} as extracted from the 
euclidean correlator shown in Fig.~\ref{fig_corr} (black and green 
lines), also compared to LO and HTL calculations, all at vanishing 
pair 3-momentum. Right figure taken from Ref.~\cite{Ding:2010ga}.}
\label{fig_drdm}
\end{figure}
It is now straightforward to convert the EM spectral functions discussed 
in the previous two sections into thermal emission rates of dileptons,
cf.~Fig.~\ref{fig_drdm}. The left panel confirms that the strong 
broadening of the $\rho$ spectral function, together with the chiral
mixing in the dip region, make the hadronic rate approach the 
partonic-based calculations, in particular the HTL-improved result.
In Ref.~\cite{Rapp:1999us} this has been interpreted as an in-medium 
reduction of the quark-hadron duality in hadronic matter with increasing 
temperature and density: the hadronic rate successively approaches the 
perturbative partonic rate, with $M_{\rm dual}\to0$ in the vicinity of
$T_c$ (or somewhat above). On the one hand, this implies an approach to 
chiral restoration, since the perturbative rate is chirally restored, 
i.e., degenerate with the axialvector. On the other hand, it also 
implies an approach to deconfinement, as signaled by the transition 
to a more economic (perturbative) description in terms of partonic 
degrees of freedom. This picture is now further corroborated by the
most recent lQCD results~\cite{Ding:2010ga}; the right panel of
Fig.~\ref{fig_drdm} exhibits a rather close agreement of the 
nonperturbative lQCD rate with the HTL rate (except at very 
small energies $q_0\ll gT $ where the latter receives additional 
corrections).   

In Ref.~\cite{Greiner:2010zg}, a ``more realistic way" to interpret 
quark-hadron duality has been suggested, by advocating the 3-volume 
multiplied production rates, $V dR/dM^2$, as the relevant quantities 
to compare. This is to account for the potentially rather different 
3-volumes of hadronic and QGP phases if considering a system at fixed 
entropy (which is a good approximation for a fireball in URHICs). 
However, the thermal production 
rates are thermodynamically intensive quantities, well-defined in the 
infinite-volume limit (in which case one should not multiply with a 
volume factor). Furthermore, as discussed in the introduction, the 
finite-$T$ QCD phase transition is most likely not a sharp one, 
especially concerning deconfinement. Recall that the increase 
in the degrees of freedom observed in the thermodynamic state
variables such as energy density is well accounted for by the 
increasing population of states in the 
hadronic resonances gas~\cite{Karsch:2003vd}. It is precisely this 
increase (a precursor of the so-called the ``Hagedorn" catastrophe) 
which leads to a likewise increase in the $\rho$ width thus signaling
its melting. In other words, the increase in degrees of freedom, which 
is ultimately responsible for a reduced volume at fixed total entropy
in a finite system, is a hadronic effect, which, in turn, is at the 
origin of the $\rho$ melting (as can be seen by glancing at the right
panel of Fig.~\ref{fig_na60-med} below).  

\begin{figure}[!t]
\begin{center}
\epsfig{file=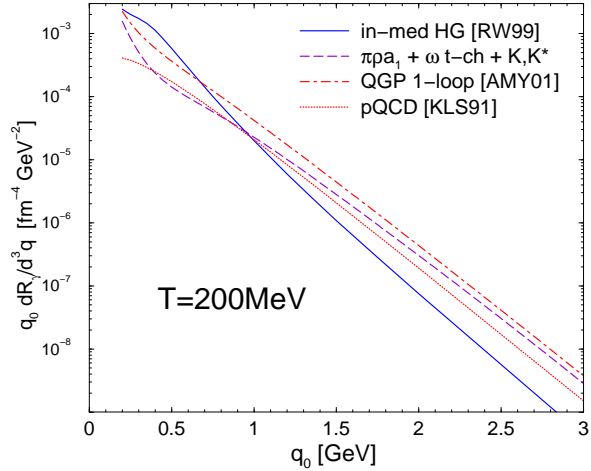,width=0.5\linewidth,angle=-90}
\end{center}
\caption{Thermal photon rates from hadronic matter~\cite{Turbide:2003si} 
(solid blue line: vector spectral function~\cite{Rapp:1999us} at the 
photon point; dashed line: $t$-channel meson-exchange reactions) and from 
the QGP (solid red line: LO HTL calculation~\cite{Kapusta:1991qp}, 
dashed-dotted line: complete resummed LO result\cite{Arnold:2001ms}).
}
\label{fig_Rgam}
\end{figure}
Thermal photon rates from the hadronic and QGP phase are compiled
in Fig.~\ref{fig_Rgam}. As elaborated following Eq.~(\ref{Rgam}),
photon production processes from a medium of on-shell particles
require a scattering process and are therefore nontrivial to
leading order in the strong coupling (e.g., $\pi\rho\to\pi\gamma$
or $qg\to q\gamma$). However, in consistently resummed many-body
calculations of the EM spectral function, the latter possesses a 
nonzero $M\to0$ limit at finite 3-momentum, see, e.g., the $\rho$
spectral function in the left panel of Fig.~\ref{fig_drho} (if
additionally $q\to 0$ one obtains the electric conductivity).  
This has been used in Ref.~\cite{Turbide:2003si} to extract the
photon production rates encoded in the hadronic spectral function
of Ref.~\cite{Rapp:1999us}, corresponding to processes of type
$\pi\rho\to a_1 \to \gamma$, $N\rho\to N^* \to N\gamma$ or pion
$t$-channel exchange in $\pi N \to \gamma N$ included in
$\Sigma_{\rho M}$, $\Sigma_{\rho B}$ and $\Sigma_{\rho\pi\pi}$,
respectively (cf.~the solid blue line in Fig.~\ref{fig_Rgam}). 
Additional $t$-channel meson-exchange reactions become important at 
photon momenta $q>1$\,GeV (cf.~the dashed line in Fig.~\ref{fig_Rgam}), 
most notably $\omega$ exchange in $\pi\rho\to\pi\gamma$. 
These rates are compared to QGP-based calculations, specifically 
to the LO HTL rate derived in Ref.~\cite{Kapusta:1991qp}, 
\begin{equation}
q_0 \frac{dR_\gamma}{d^3q} = \frac{6}{9} \frac{\alpha\alpha_S}{2\pi^2}
       T^2 {\rm e}^{-q_0/T}
       \ln\left(1+\frac{2.912}{4\pi\alpha_s} \frac{q_0}{T} \right) \  ,
\label{rate_pQ}
\end{equation}
for $N_f$=3 light flavors and with an added ``1" in the logarithm to 
regularize its infrared behavior (surprisingly,
the such obtained electric conductivity is not far from the one 
computed in thermal lQCD discussed in Sec.~\ref{ssec_qgp}).
This expression nicely illustrates the leading ${\cal O}(\alpha_s\alpha)$
behavior of the photon rate. It turns out, however, that there
are additional contributions at this order, requiring a full resummation
of ladder diagrams. This has been achieved in Ref.~\cite{Arnold:2001ms},
The numerical result of this work exceeds the ``naive" LO rate
by about a factor of 2-3, cf.~Fig.~\ref{fig_Rgam}. 
When comparing QGP and total hadronic rates (the sum of the
two contributions plotted in Fig.~\ref{fig_Rgam}), one again finds
that both are very comparable at a temperature of $T$=200\,MeV (chosen 
mostly for historic reasons). Thus photon rates also
support the duality hypothesis, although the theoretical control
over the calculations is less good as in the dilepton case.   

\section{Dilepton and Photon Spectra in Heavy-Ion Collisions}
\label{sec_hics}
To date, dilepton spectra in heavy-ion collisions have been measured at 
the SPS ($\sqrt{s}$=17.3,8.7\,AGeV)~\cite{Agakichiev:2005ai,Adamova:2003kf,Arnaldi:2006jq,Adamova:2006nu,Arnaldi:2008er,Damjanovic:2009zz} 
at RHIC ($\sqrt{s}$=200\,AGeV) \cite{Adare:2009qk,Zhao:2011wa} and at lower 
energies at BEVALAC and SIS 
($E_{\rm lab}$=1-2\,AGeV)~\cite{Porter:1997rc,Agakishiev:2007ts}, see, 
e.g., Refs.~\cite{Tserruya:2009zt,Specht:2010xu} for recent experimental 
reviews. In this section we apply the rates discussed above to calculate
thermal-emission spectra from URHICs, focusing our comparison on recent
highlights from data at SPS and RHIC. Two additional ingredients 
are required to do so: a realistic space-time evolution of the expanding 
fireball formed in the reaction and an assessment of non-thermal 
sources contributing to the thermal spectra (in some cases it
is possible to subtract these experimentally).  

Basic elements of a realistic fireball are summarized in
Sec.~\ref{ssec_fb}, while non-thermal sources are addressed on a 
case-by-case bases of each observable. The analysis of experimental 
spectra is organized according to their putative information content, 
namely in-medium effects on spectral properties in connection with 
the emission duration (Sec.~\ref{ssec_spec}) and temperature and 
expansion characteristics of the source (Sec.~\ref{ssec_thermo}). 

\subsection{Space-Time Evolution of Heavy-Ion Collisions}
\label{ssec_fb}
\begin{figure}[!t]
\begin{center}
\epsfig{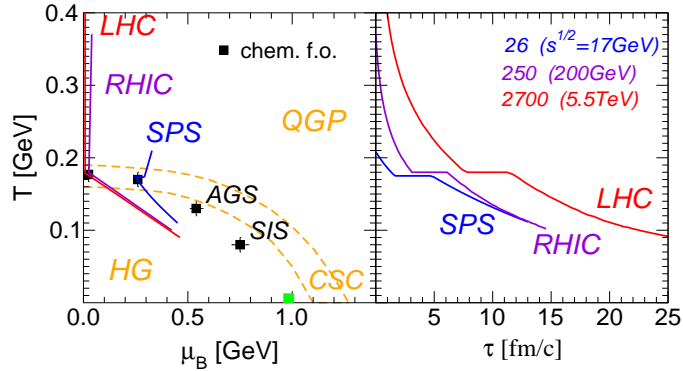}
\end{center}
\caption{Schematic representation of isentropic trajectories that 
the medium follows in URHICs at SPS, RHIC and LHC in the QCD phase 
diagram (left panel) and when projected on the temperature axis as a 
function of time using hadron spectra to estimate the collective 
expansion properties (right panel). Each trajectory is calculated for a 
fixed specific entropy, $S/N_B=s/\varrho_B$, according to entropy and 
baryon-number conservation. In addition, the hadronic part of the 
trajectories implements effective chemical potentials for all strongly 
stable hadrons~\cite{Rapp:2002fc} to maintain the observed hadron 
ratios as determined by thermal-model fits at chemical 
freezeout~\cite{BraunMunzinger:2003zd} characterized by the ``data 
points".}
\label{fig_evo}
\end{figure}

Much has been learned about the space-time evolution of the medium 
formed in URHICs from the wealth of available hadron abundances, $p_t$ 
spectra and elliptic flow. Generally, hydrodynamic models do 
well in describing the latter two (especially at RHIC and LHC) up
to momenta of $p_t$=2-3\,GeV. This encompasses more than 90\% 
of the observed particles, and justifies the notion of a collectively
expanding medium in local thermal equilibrium where the pressure
gradients drive the acceleration. This mechanism works well enough to 
have predicted azimuthal asymmetries in the $p_t$ spectra 
for non-central heavy-ion collisions: the elliptic asymmetry given by 
the transverse overlap zone 
of the colliding nuclei leads to larger pressure gradients along the 
short axis of the ellipsoid, which in turn provide larger acceleration 
resulting in a positive ``elliptic flow" coefficient, $v_2(p_t)$, in 
the angular distribution of the hadron $p_t$ spectra (this is arguably 
the strongest evidence for the celebrated ``near-perfect liquid" 
discovered at RHIC). Taking all evidence together, the following 
picture emerges: Shortly after the collision of the two nuclei, at a 
thermalization time of $\tau_0$=0.5-1\,fm/$c$, most of the entropy
has been produced and the system has thermalized into a QGP, with 
initial temperature of ca.~$T_0$=350-400\,MeV (200-250\,MeV at SPS, 
$\sim$500-600\,MeV at LHC). The QGP expands for about 3\,fm/$c$ before 
entering the transition regime to hadronic matter at around 
$T_c\simeq170$\,MeV, where it spends almost 5\,fm/$c$. At this point 
the hadronic abundances become frozen, leading to the notion 
of ``chemical freezeout" where the inelastic reactions are believed to 
turn off. In the subsequent hadronic phase, elastic scattering driven
by hadronic resonance is still operative leading to a further cooling 
of the system for about 5-10\,fm/$c$ until ``thermal freezeout" 
occurs at $T_{\rm fo}\simeq 100$\,MeV.
After this, hadrons stream freely to the detector (modulo long-lived
electromagnetic and weak decays).   
Figure~\ref{fig_evo} summarizes typical trajectories in the phase 
diagram, as well as the time dependence of temperature, for central 
heavy-ion collisions at SPS, RHIC and LHC.
Besides initial conditions and the equation of state, hydrodynamic 
models require the implementation of ``freezeout". At thermal freezeout, 
the medium is converted into hadron spectra using the time-honored
Cooper-Fry prescription. ``Chemical freezeout" implies that in the
subsequent evolution the abundances of observed (stable) hadrons should 
be conserved. This is achieved by introducing effective chemical
potentials into the hadronic equation of state (usually modeled via
a hadron resonance gas), to individually conserve the number of pions, 
kaons, nucleons and antinucleons (not only their difference, which is an
exact conservation law related to the baryon chemical potential,
$\mu_B$), etc.~\cite{Rapp:2002fc}. For dilepton (photon) production 
this implies that, e.g., the process $\pi\pi\to\rho\to e^+e^-$  
($\pi\rho\to\pi\gamma$) is augmented by a pion fugacity squared, 
$z_\pi^2$ ($z_\pi^3$) where $z_\pi=\exp(\mu_\pi/T)$.

\begin{figure}[!t]
\begin{center}
\epsfig{file=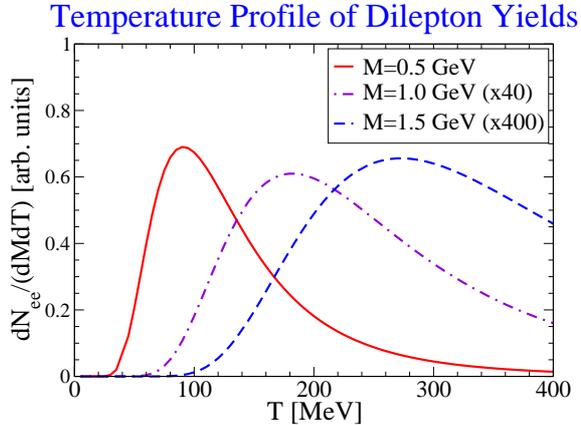,width=0.5\linewidth,angle=-90}
\end{center}
\caption{Emission yields of thermal dileptons as a function of
temperature for different values of their invariant mass, assuming
a temperature independent EM spectral function.
}
\label{fig_dndT}
\end{figure}

Before going into quantitative theory comparisons to data, it is
instructive to make a more general assessment of which stages of the 
fireball evolution will contribute at which dilepton energies/masses in 
the spectra. Toward this goal we write the the time-differential 
invariant-mass spectrum, integrated over 3-momentum and 3-volume
at each time snapshot, as~\cite{Rapp:2004zh}
\begin{eqnarray}
\frac{dN_{ee}}{dMd\tau} &=&
\frac{M}{q_0} \int d^3x \ d^3q \ \frac{dN_{ee}}{d^4xd^4q}
\nonumber\\
 &\simeq & {\rm const} \  V_{FB}(\tau) \ \frac{{\rm Im}\Pi_{\rm em}(M;T)}{M}
           \int \frac{d^3q}{q_0} \ {\e}^{-q_0/T}
\nonumber\\
 &\simeq & {\rm const} \  V_{FB}(\tau) \ \frac{{\rm Im}\Pi_{\rm em}(M;T)}{M^2}
           \ \e^{-M/T} \ (MT)^{3/2} \ , 
\end{eqnarray}
where $V_{\rm FB}$ denotes the volume of the expanding fireball. For the 
second (approximate) equality it has been assumed that the in-medium EM 
spectral function depends only weakly on 3-momentum (which is exact in 
vacuum due to Lorentz-invariance, and approximately satisfied in medium 
for not too large momenta). For the third (approximate) equality we have 
invoked the non-relativistic approximation, $T/M\ll1$. In a last step, 
we relate the fireball-evolution time to the temperature, $T(\tau)$, 
to obtain the temperature-differential emission as
\begin{equation} 
\frac{dN_{ee}}{dMdT}\propto  
{\rm Im}\,\Pi_{\rm EM}(M;T) \ {\rm e}^{-M/T}  \ T^{-m} \ . 
\end{equation}
The power of temperature, $m$, depends on the details of the fireball
expansion (e.g., 1-D Bjorken or full 3-D as appropriate for the early 
and late phases, respectively), and the equation of state
(i.e., the power law of entropy density, $s\propto T^n$, with $n$=3 for
an ideal massless gas, but larger for a hadron-resonance gas), see
Ref.~\cite{Rapp:2004zh} for more details. 
In Fig.~\ref{fig_dndT} we plot the coefficient $f(T;M)=\exp(-M/T) \ 
T^{-m}$ as a function of temperature for different masses. At each 
mass, the maximal emission yields are centered around a temperature 
$T_{\rm max}\simeq T/m \simeq T/5$ which  reflects the competition
between the thermal exponential, favoring high temperature, and
the increasing 3-volume of the fireball as it expands and cools.   
Note that for low-mass dilepton emission, $M\le 1$\,GeV, the profile 
function, $f(T;M)$, peaks in the hadronic phase of the evolution. On 
the one hand, medium effects in the EM spectral function, which lead to 
a marked enhancement for $M<m_\rho$ with increasing $T$ and $\varrho_B$
(recall Fig.~\ref{fig_drho}), will shift this peak to somewhat 
larger $T$ (note, however, that the low-mass enhancement tends
to saturate when approaching $T_c$, see Fig.~\ref{fig_drdm} left). 
On the other hand, the latent heat burned off in the hadron-to-parton
transition, i.e., the rapid rise of $s/T^3$ in the vicinity of the 
phase conversion, will bias the emission to the hadronic phase. Both
effects lead to an additional focusing of low-mass dilepton emission
to hot and dense hadronic matter. 

Similar arguments apply to thermal photon (or dilepton) 
transverse-momentum spectra, with an important amendment due to
the transverse flow of the fireball causing a spectral blue shift
(well-known from hadron spectra). Since the flow develops over time,
it creates a bias toward the later emission stages.

\subsection{Spectrometer and Chronometer}
\label{ssec_spec}
\begin{figure}[!t]
\begin{minipage}{0.5\linewidth}
\hspace{-0.3cm}
\epsfig{file=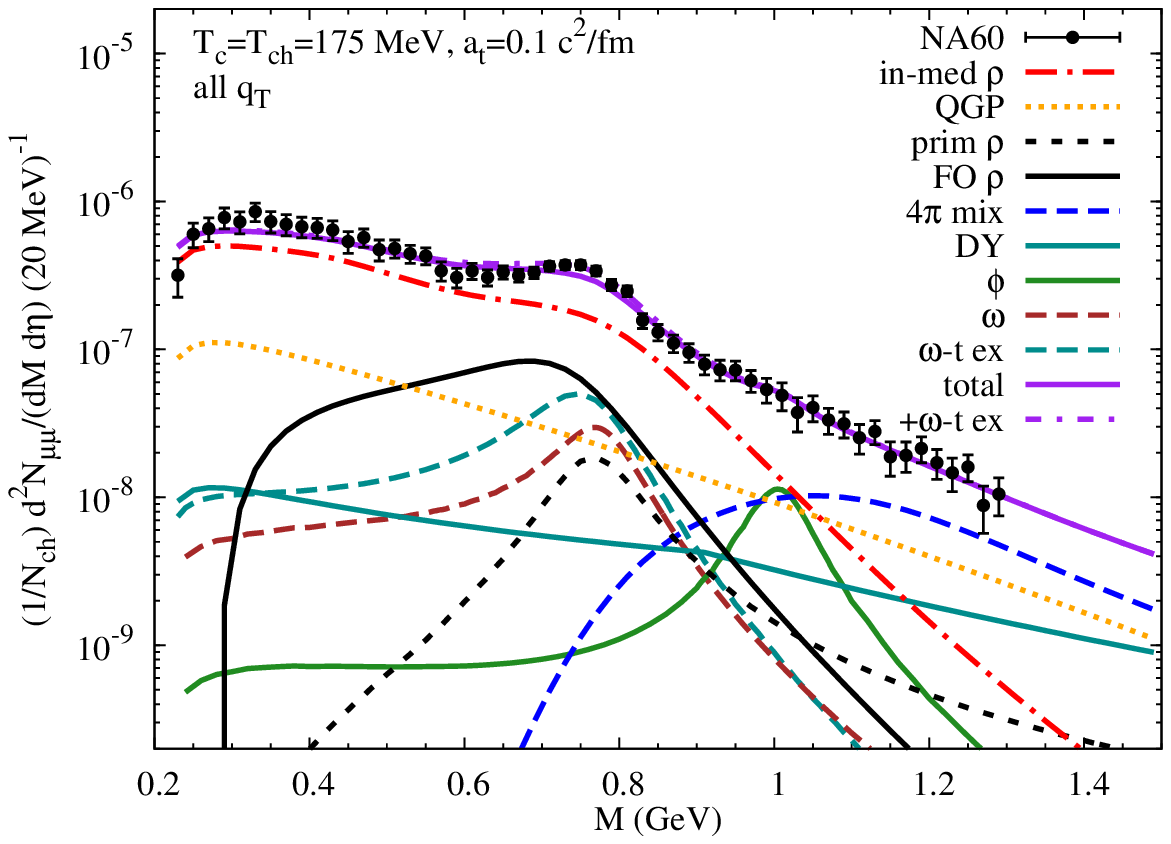,width=0.94\linewidth}

\epsfig{file=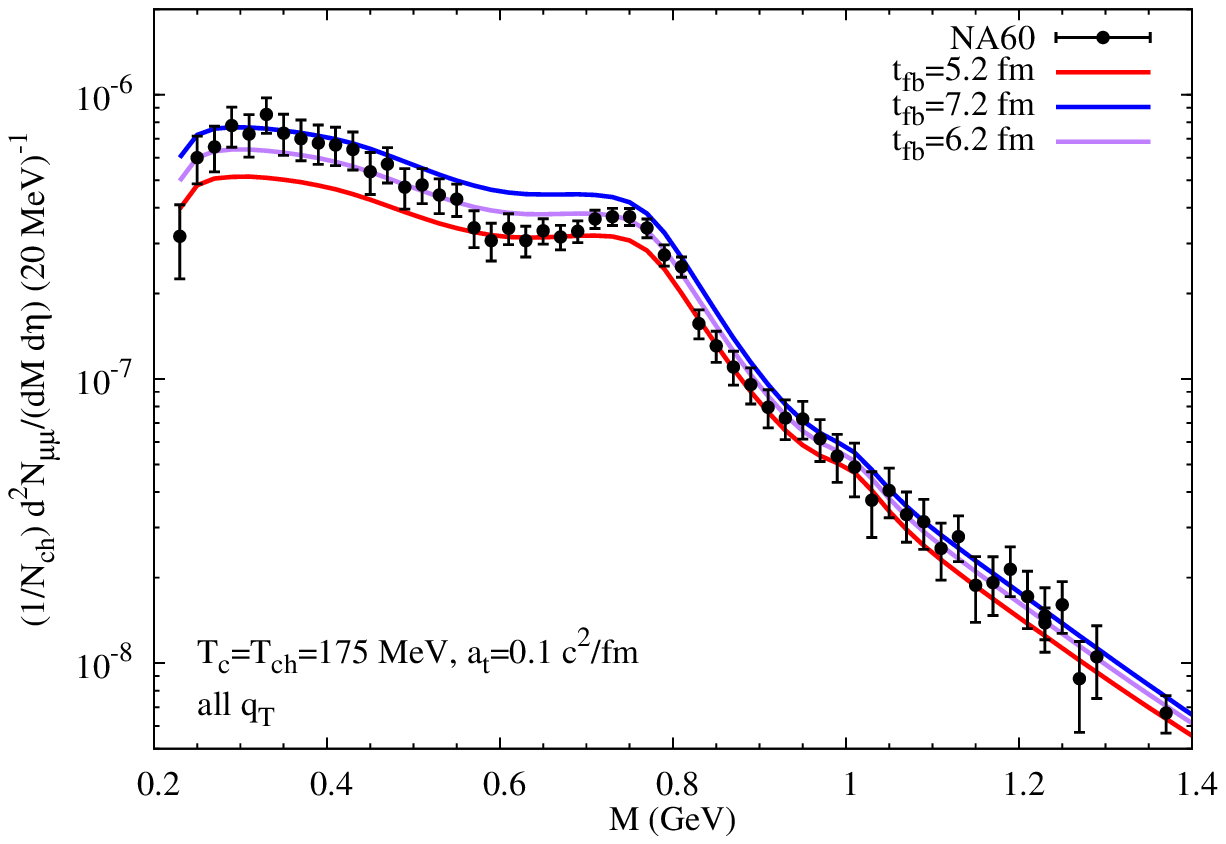,width=0.94\linewidth}
\end{minipage}
\begin{minipage}{0.5\linewidth}
\vspace{0.0cm}
\epsfig{file=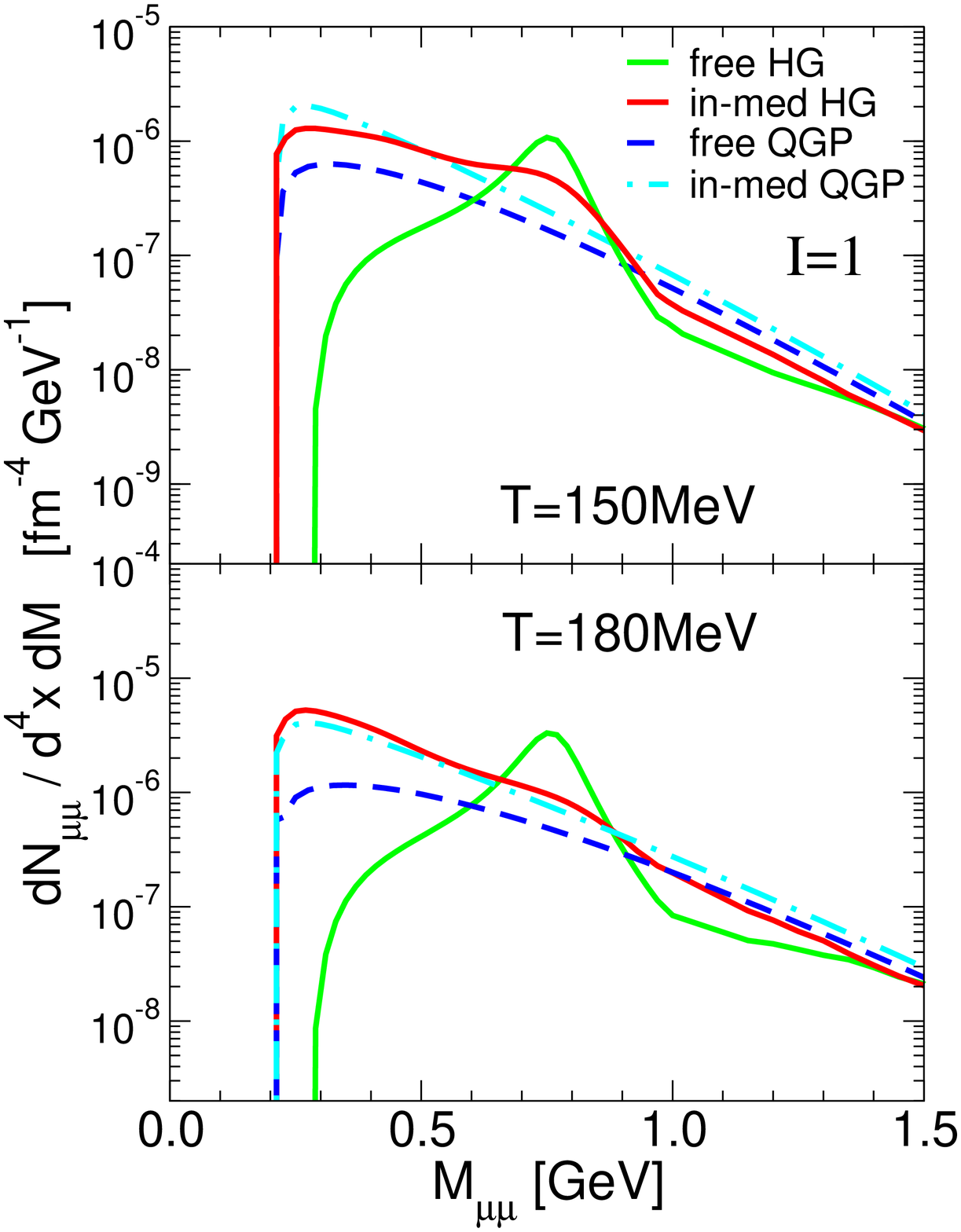,width=0.97\linewidth}
\end{minipage}
\vspace{0.1cm} 
\caption{Left panels: excess dimuon invariant-mass spectra as measured 
in In-In($\sqrt{s}$=17.3\,AGeV) collisions at the
SPS~\cite{Arnaldi:2008er,Damjanovic:2009zz}, compared to theory 
calculations with in-medium hadronic spectral functions and QGP
emission from an expanding fireball~\cite{vanHees:2007th}. The upper 
panel resolves the
individual contributions in the calculation, which, however are 
dominated by the in-medium $\rho$ spectral function at low-mass
and a combination of hadronic continuum (``4$\pi$") and QGP radiation 
at intermediate mass. The in-medium $\rho$ plus 4$\pi$ make up the 
total in-medium hadron-gas (HG) rate shown by the solid red line in the 
right panels for two temperatures (including baryon-density 
effects)~\cite{Rapp:1999ej}.  The lower left panel illustrates the 
effect of varying the fireball lifetime in the calculation.}
\label{fig_na60-M}
\end{figure}

The state-of-the art in dilepton measurements in URHICs has been 
set by the NA60 collaboration with their dimuon data from 
In-In($\sqrt{s}$=17.3\,AGeV) collisions at the 
SPS~\cite{Arnaldi:2006jq,Arnaldi:2008er}. Excellent statistics combined 
with superb mass resolution allowed for a subtraction of background 
sources thus isolating the ``excess radiation" from reinteractions in 
the fireball. In addition, a full acceptance correction could be carried 
out so that the invariant-mass spectra, displayed in the left panels of
Fig.~\ref{fig_na60-M}, are, for the first time, truly {\em invariant}. 
This, in particular, implies that blue-shift effects of the expanding
fireball do not distort the spectra. The result is quite stunning:
the data show a near-perfect thermal spectrum, ``only" modulated by a 
broad bump around the free $\rho$/$\omega$ masses. Even without any
fireball convolution, a comparison to the theoretical input rates 
calculated in Ref.~\cite{Rapp:1999ej} (shown in the left panel of 
Fig.~\ref{fig_drdm}, supplemented with dimuon threshold kinematics)
exhibits a remarkable agreement for an average temperature of 
150-160\,MeV. This temperature is tantalizingly close to the most 
recent lQCD predictions shown in the left panel of Fig.~\ref{fig_lat}, 
provoking the conclusion that the chiral transition has been observed at 
SPS! Of course, these rather general arguments need to be backed up by 
quantitative calculations convoluting the $T$- and $\varrho_B$-dependent 
emission rates over a realistic fireball expansion (as discussed in 
Sec.~\ref{ssec_fb}). This has been done in Ref.~\cite{vanHees:2007th}
and the results are in very good agreement with the data as seen in
the upper left panel of Fig.~\ref{fig_na60-M}. Subleading contributions 
include the decays of $\omega$ and $\phi$ mesons, a non-thermal 
component of $\rho$ mesons, and Drell-Yan annihilation. 
Especially the latter two play a significant role when high-$q_t$
cuts are imposed on the spectra (this is fully in line with the
standard notion from hadron $p_t$ spectra that the thermal component 
dominates up to $p_t\simeq$2\,GeV yielding to ``hard production"
above). With the level of precision set by the NA60 mass spectra
several further diagnostics of the produced medium become available,
as discussed in the following.    

\begin{figure}[!t]
\begin{minipage}{0.5\linewidth}
\epsfig{file=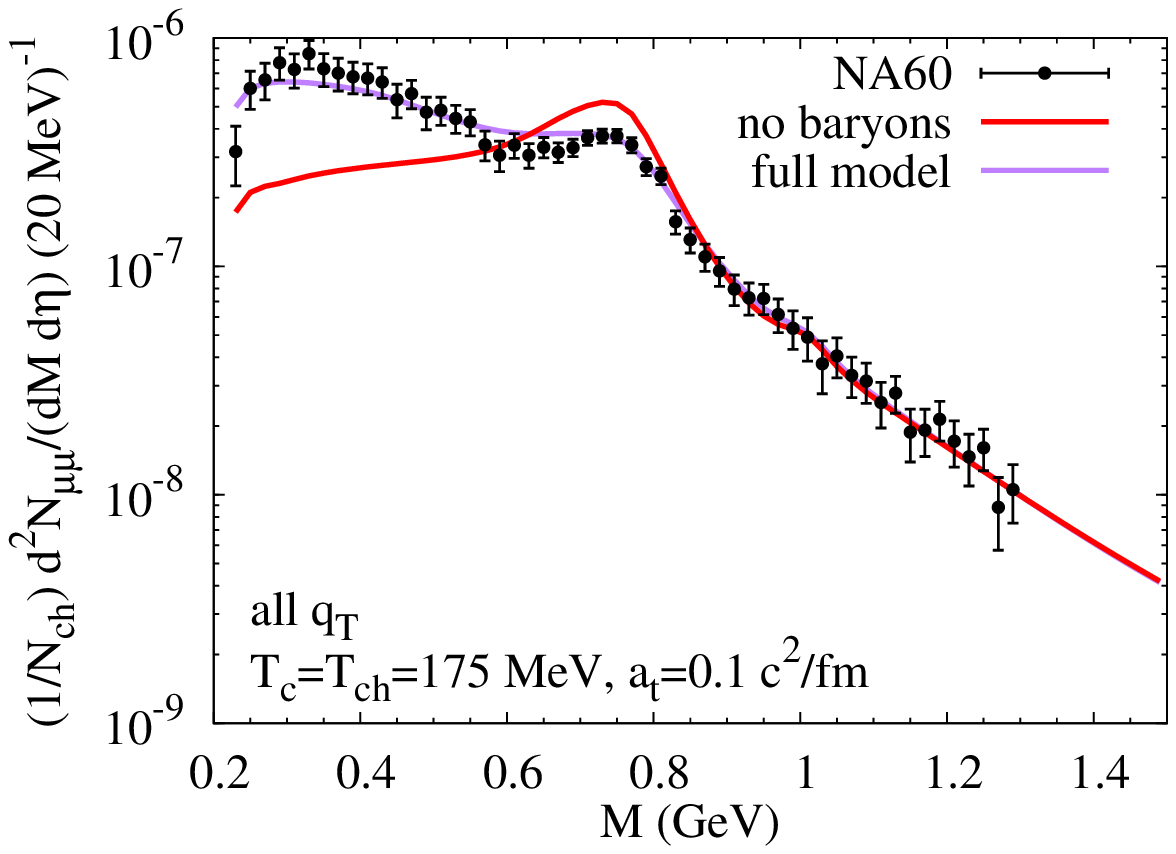,width=0.97\linewidth}
\end{minipage}
\begin{minipage}{0.5\linewidth}
\vspace{-0.6cm}
\epsfig{file=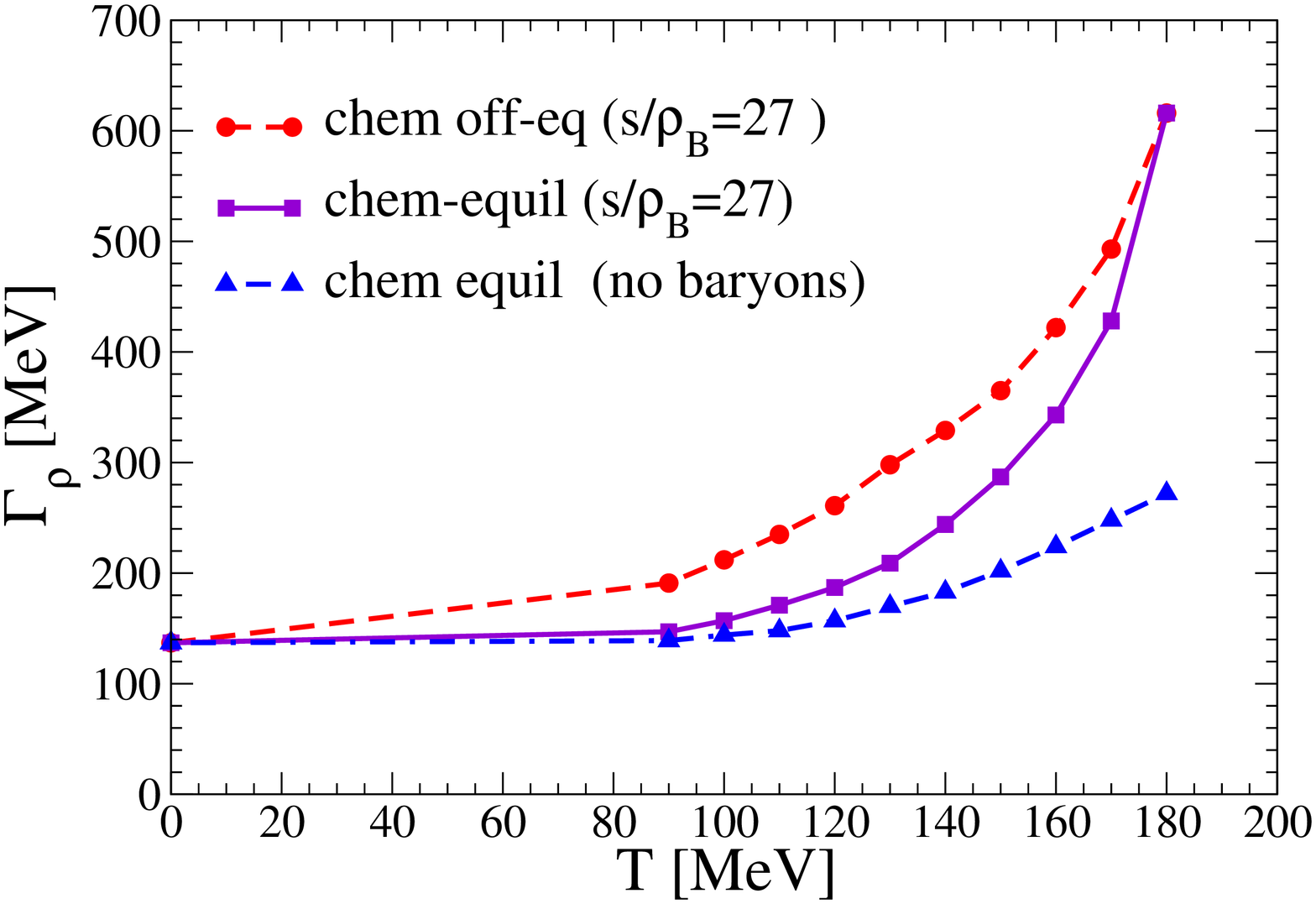,width=0.84\linewidth,angle=-90}
\end{minipage}
\vspace{0.1cm}
\caption{Left panel: comparison of the acceptance-corrected NA60 mass
spectra in In-In($\sqrt{s}$=17.3\,AGeV)
collisions~\cite{Arnaldi:2008er,Damjanovic:2009zz} to a calculation
where baryon-induced effects ion the EM spectral function have been
switched off. Right panel: in-medium $\rho$ widths (estimated
from the full-width-half-maximum of its spectral function) along
the SPS trajectory used in the spectra calculations projected on
the temperature axis, for the full calculation including effective
chemical potentials (circles) and in chemical equilibrium  with
(squares) and without (triangles) baryon-induced effects; figure
taken from Ref.~\cite{Rapp:2009yu}.
}
\label{fig_na60-med}
\end{figure}
The first obvious question is how sensitive the mass spectra are to 
the $\rho$-meson spectral shape. This has been tested qualitatively by 
switching off the medium effects induced by baryons, i.e., using the 
$T$-dependent $\rho$ spectral function in a hot meson gas as displayed
in the right panel of Fig.~\ref{fig_drho}. The resulting $\mu^+\mu^-$
spectra are in stark disagreement with the NA60 data, cf.~left panel
of Fig.~\ref{fig_na60-med}: the low-mass enhancement is underestimated
by about a factor of $\sim$4, and the yield in the $\rho$-peak region
is overestimated by about a factor of $\sim$2, both several standard 
deviations away from experiment. The evolution of the $\rho$ width
(taken as full-width-half-maximum in the spectral function) along the 
SPS trajectory used in the calculations is displayed in the right panel 
of Fig.~\ref{fig_na60-med}. The upper curve (circles), corresponding to 
the full calculation, confirms that the ``average width" of 
$\bar\Gamma_\rho^{\rm med}\simeq$\,350-400\,MeV reflects the conditions
of the medium close to the chiral transition. It also exhibits a rather
strong rise for temperatures beyond that. This rise becomes even more 
pronounced in chemically equilibrated matter where the effective 
meson-chemical potentials are switched off (squares; at fixed 
$s/\varrho_B$ this also implies a slight reduction in $\mu_B$). It is 
not unlike the increase of the thermodynamic state variables in the 
hadron-resonance gas around a similar temperature. No such behavior 
emerges at comparable temperatures if baryon-induced medium effects are 
switched off. Since the average width implies the occurrence of larger 
widths in the course of the medium expansion, a situation with 
$\Gamma_\rho^{\rm med}\simeq m_\rho$ must be realized, i.e., resonance 
melting.
 
It is important to note that baryon-induced medium effects remain a key 
player at RHIC~\cite{Rapp:2000pe}. This is so because the $\rho$ equally 
interacts with baryons ($B$) and anti-baryons ($\bar B$) due to CP 
invariance of strong interactions (the $\rho$ is a CP eigenstate). 
Therefore, the sum (not the difference) of $\varrho_B$ and 
$\varrho_{\bar B}$ is the relevant quantity for medium effects. At RHIC, 
the rapidity density of baryons plus antibaryons is slightly larger than 
at SPS, while the pion rapidity density is roughly a factor of 2 larger,
implying an accordingly increased fireball volume at given temperature.
Thus, the ``effective" density 
$\rho_{\rm eff}\equiv \varrho_B + \varrho_{\bar B}$, is not very 
different at SPS and RHIC, and probably also at LHC. From a theoretical 
point of view, the configuration of small net-baryon
chemical potential at collider energies is appealing because it comes
closest to $\mu_B$=0 where lQCD computations are most powerful.    

While the spectral shape of the NA60 data could be predicted by theory,
the total yields could not, at least not within the experimental 
accuracy. However, one can turn the argument around and utilize the 
total yield as a precision clock for the emission duration. In previous 
analyses of dilepton spectra in Pb-Au collisions by 
CERES~\cite{Agakichiev:2005ai}, the fireball lifetime estimated from
hydrodynamic models, $\tau_{\rm FB}\simeq10-15$\,fm/$c$, resulted in
a reasonable description of the $e^+e^-$ excess
radiation~\cite{Rapp:1999us}. The precision attained with the NA60
data is significantly increased: even a variation of $\pm1$fm/$c$
in the calculation discussed above leads to noticeable deviations
from the data, cf.~lower left panel of Fig.~\ref{fig_na60-M}. The 
resulting ``measurement" of $\tau_{rm FB}=6.5\pm1$fm/$c$ for the 
fireball lifetime using dilepton yields is probably the most accurate 
one thus far in URHICs (it is not to be confused with the pion 
emission duration referred to in Hanbury-Brown Twiss (HBT) 
interferometry measurements).  

Let us briefly comment on the current situation of low-mass
dilepton measurements at RHIC. The PHENIX collaboration found a very
large enhancement for masses below the free $\rho$ mass in central 
Au-Au($\sqrt{s}$=200\,AGeV) collisions~\cite{Adare:2009qk} which
is largely concentrated at low transverse momenta, $q_t<0.5$\,GeV.
It cannot be explained by the theoretical ingredients which allow
for an understanding of the SPS dileptons.  
The STAR collaboration, on the other hand, reports a smaller
enhancement~\cite{Zhao:2011wa}, which, however, is still sizable
and roughly consistent with theoretical expectations.

\subsection{Thermometer and Barometer}
\label{ssec_thermo}
\begin{figure}[!t]
\begin{minipage}{0.5\linewidth}
\epsfig{file=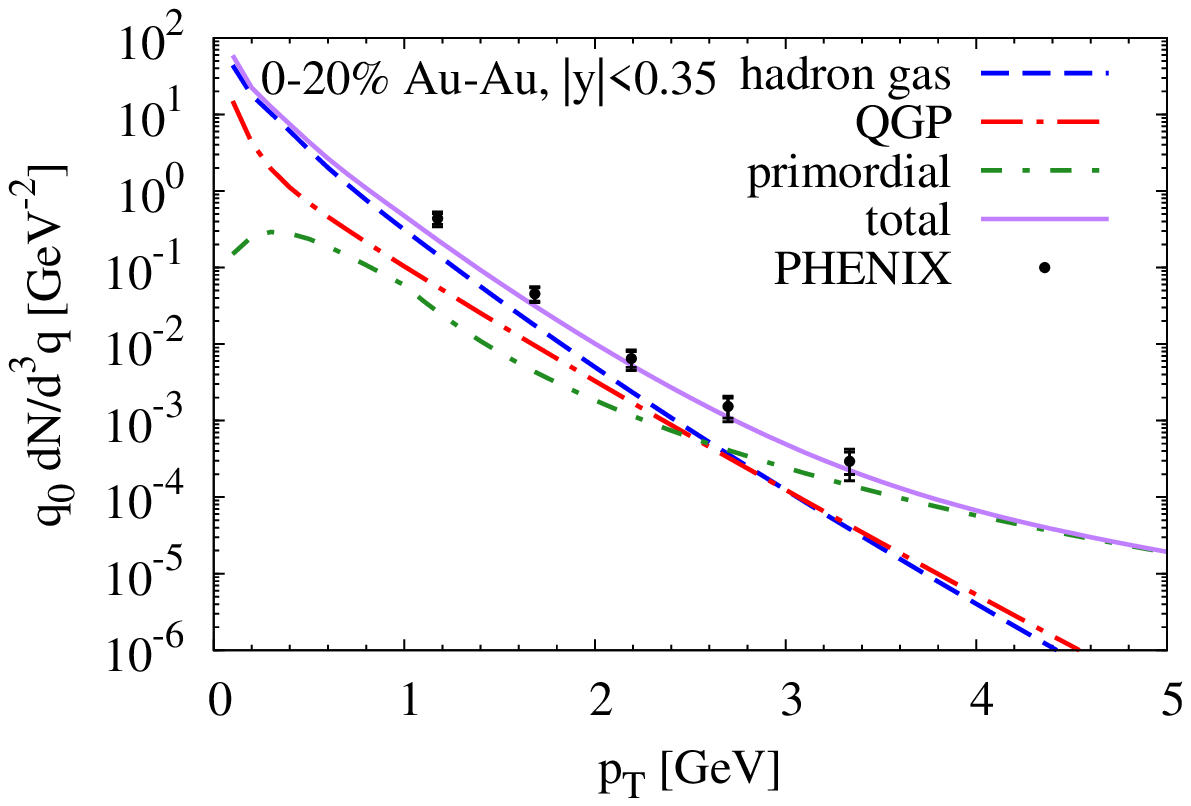,width=0.98\linewidth}
\end{minipage}
\begin{minipage}{0.5\linewidth}
\epsfig{file=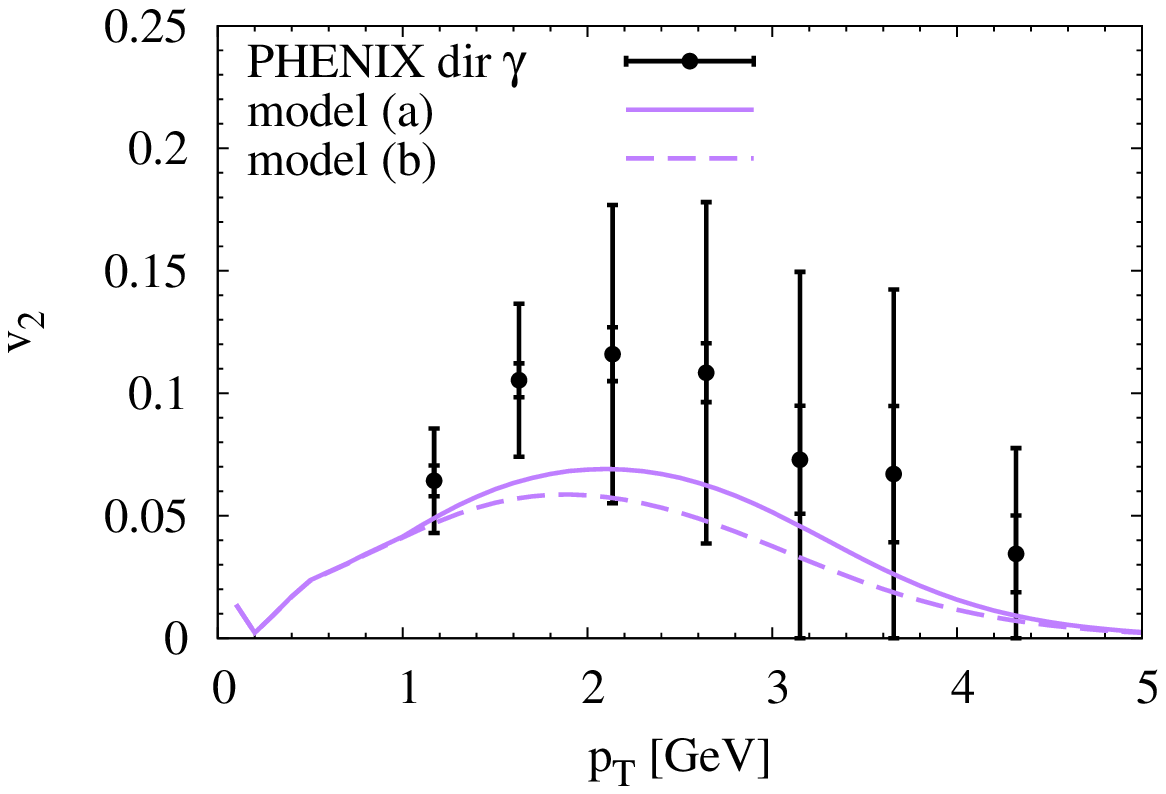,width=0.98\linewidth,angle=-0}
\end{minipage}
\vspace{0.1cm}
\caption{Transverse-momentum spectra (left panel) and elliptic flow 
(right panel) of direct photons in 0-20\% central 
Au-Au($\sqrt{s}$=200\,AGeV) collisions at RHIC, compared to PHENIX
data~\cite{Adare:2008fqa,Adare:2011zr}. The curves are 
calculations~\cite{vanHees:2011vb} with a realistic fireball evolution 
employing thermal QGP and hadronic rates which are ``dual" around 
$T_c$, corresponding to Fig.~\ref{fig_Rgam}.}
\label{fig_phenix-gam}
\end{figure}

Historically, the best experimental tool to serve as a ``thermometer" of 
the early phases in URHICs was believed to be thermal radiation of photons.
With a structureless emission rate, the slope of the spectrum is directly
related to the temperature of the system, modulo a blue-shift from 
transverse expansion which is expected to be small in the early stages. 
At SPS energies, the presence of a significant excess signal of photons, 
i.e., beyond those from initial production and final-state decays of 
long-lived hadrons (mostly $\pi^0,\eta\to \gamma\gamma$), remains 
inconclusive~\cite{Aggarwal:2011ns}.   
At RHIC, however, the PHENIX collaboration has observed a strong excess
signal in semi-/central Au-Au collisions~\cite{Adare:2008fqa} for 
transverse momenta above $q_t=1$\,GeV up to at least 3-4\,GeV, beyond 
which the primordial production extrapolated from binary $N$-$N$ 
collisions dominates, see left panel of Fig.~\ref{fig_phenix-gam} 
(``excess" plus ``primordial" are usually referred to as ``direct" 
photons, i.e. long-lived decays subtracted). This range nicely coincides 
with theoretical expectations of where the QGP radiation is most 
visible~\cite{Alam:1999sc,Turbide:2003si,Chatterjee:2005de,Liu:2009kta,Holopainen:2011pd,Dion:2011pp}. 
The inverse slope of the excess radiation was found to be consistent
with an exponential with temperature $T=221\pm19^{\rm stat}
\pm19^{\rm sys}$\,MeV. This seems to point at a source well inside
the QGP, albeit not to the very early phases where temperatures above 
300\,MeV are expected, recall~Fig.~\ref{fig_evo}.  

Maybe even more intriguing are the recent PHENIX measurements of the
direct-photon elliptic flow~\cite{Adare:2011zr}, cf.~right panel of
Fig.~\ref{fig_phenix-gam}. In the regime of the excess radiation, i.e.,
for $q_t\le 3$\,GeV, the elliptic-flow coefficient, $v_2(q_t)$, is 
comparable to those of pions, which are only emitted at the end of the 
fireball evolution, i.e., with maximal $v_2$. On the contrary, the 
excess photons are supposed to be emitted throughout the evolution
(including early, where the $v_2$ is still small), and the primordial 
ones should carry zero $v_2$. More quantitatively, predictions of 
thermal-photon $v_2$ using hydrodynamic models with QGP-dominated 
excess radiation lead to maximal values of $v_2\simeq2$\% for direct 
photons in 0-20\% Au-Au($\sqrt{s}$=200\,AGeV) 
collisions~\cite{Chatterjee:2005de,Liu:2009kta,Holopainen:2011pd,Dion:2011pp}, 
well below the PHENIX data. In these calculations the hadronic emission
rates were significantly smaller than the QGP rate in the vicinity of 
$T_c$, which differs from the ``duality" hypothesis suggested by 
Fig.~\ref{fig_Rgam}. In Ref.~\cite{vanHees:2011vb} we therefore revisited 
our earlier calculations of direct photons~\cite{Turbide:2003si} by 
refining the thermal fireball model as to quantitatively describe the 
$p_t$ spectra and elliptic flow of bulk hadrons, freezing out at 
$T_{\rm fo}\simeq100$\,MeV, and multistrange hadrons, kinetically
decoupling close to chemical freezeout, $T_{\rm ch}\simeq 180$\,MeV.  	
This, in particular, led to harder thermal photon spectra from the
hadronic phase, thus exceeding the QGP contribution over the for
thermal emission relevant $q_t$ range up to 3-4\,GeV. Consequently,
the thermal-photon $v_2$ increases substantially, now being dominated
by hadronic emission where most of the final $v_2$ has already been
built up. The resulting direct-photon $v_2(q_t)$, which includes
the primordial component (extrapolated from $pp$ collisions) with
vanishing $v_2$, reaches up to 6-7\%, a factor of $\sim$3 larger than 
in previous calculations and thus much closer to experiment, cf.~right
panel in Fig.~\ref{fig_phenix-gam}.

\begin{figure}[!t]
\begin{minipage}{0.5\linewidth}
\epsfig{file=phot-slope.eps,width=0.96\linewidth}
\end{minipage}
\begin{minipage}{0.5\linewidth}
\epsfig{file=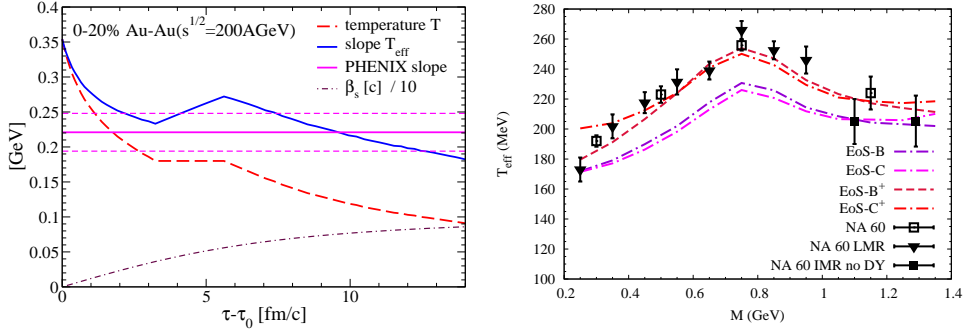,width=0.98\linewidth,angle=-0}
\end{minipage}
\vspace{0.1cm}
\caption{Effective slope parameters of transverse-momentum spectra 
of EM radiation in URHICs. Left panel: time dependence of $T_{\rm eff}$
of thermal photons, Eq.~(\ref{Teff}), in an expanding fireball for 
0-20\% central Au-Au($\sqrt{s}$=200\,AGeV) collisions at RHIC (solid 
blue line), compared the local temperature (dashed line) and PHENIX 
data~\cite{Adare:2011zr} (horizontal solid line with dashed-line 
error band); the dash-dotted line indicated the flow velocity of
the fireball surface. Right panel: $T_{\rm eff}$ for excess dimuons
as a function of their invariant mass as measured by NA60 in 
In-In($\sqrt{s}$=17.3\,AGeV) collisions at SPS~\cite{Arnaldi:2008er},
compared to theory calculations~\cite{vanHees:2007th} with different 
transverse fireball 
acceleration (lower 2 curves: $a_t=$0.085\,fm/$c^2$, upper 2 curves:
$a_t=$0.1\,fm/$c^2$) and with different phase-transition
temperatures (EoS-B: $T_c$=160\,MeV, EoS-C: $T_c$=190\,MeV).
}
\label{fig_slopes}
\end{figure}
With dominantly hadronic emission, one should revisit the evaluation of
the photon slope parameter. For this purpose, we plot in the left panel 
of Fig.~\ref{fig_slopes} the blue-shifted effective temperature as a 
function of time in the underlying fireball evolution,
\begin{equation}
T_{\rm eff} \simeq T \sqrt{\frac{1+\langle\beta\rangle}
{1-\langle\beta\rangle}} \ ,
\label{Teff}
\end{equation}
where $T$ is the ``true" temperature in the thermal rest frame and
$\langle\beta\rangle$ the average transverse flow velocity (typically 
2/3 of the surface velocity). The comparison to the PHENIX measurement 
reveals that the largest overlap is indeed inside the hadronic phase, 
for a broad temperature interval around $\sim$130\,MeV, 
``Doppler-shifted" by about 100\,MeV.
This is consistent with the large elliptic flow.  

Similar analysis can be carried out for dileptons, which has been done 
with the NA60 dimuon $q_t$-spectra at the SPS~\cite{Arnaldi:2008er}. The 
extracted slope parameters are displayed as a function of the 
invariant-mass window imposed on the $q_t$ spectra in the right panel 
of Fig.~\ref{fig_slopes}. One finds a continuous rise of $T_{\rm eff}$
with mass, roughly following the schematic (nonrelativistic) analogue
of Eq.~(\ref{Teff}) for massive particles (here: virtual photons),
\begin{equation}
T_{\rm eff} \simeq T + M \langle\beta\rangle^2 \ . 
\label{Teff2}
\end{equation}
This corroborates the hadronic nature of low-mass dileptons below the
$\rho$ mass. However, above the $\rho$ mass, the slopes decrease,
which has been argued to signal a transition to a new source with
small flow and thus of partonic nature. However, the latter part
of the interpretation is not unique and one should keep in mind
the following features: (a) the $\rho$-mass region, 
$M\simeq0.7$-$0.9$\,GeV, is predominantly populated by contributions 
from the late fireball stages, since the $\rho$ spectral shape is 
recovering its vacuum form which prominently
figures around $M\simeq m_\rho$; (b) the interplay of thermal factor
and 3-volume implies that going up in mass the emission is ``biased" 
toward the earlier phases (recall Fig.~\ref{fig_evo}), which increases
$T$ but more strongly decreases the Doppler shift in Eq.~(\ref{Teff2}),
$M\langle\beta\rangle^2$;
(c) for temperatures around $T_c$ the duality of the emission 
rates renders a ``partonic" and ``hadronic" assignment ambiguous,
depending on the not-so-well-defined definition of the transition
temperature. The latter is demonstrated by the two pair of lines
in Fig.~\ref{fig_slopes}, where each pair is calculated assuming two
``extreme" values of $T_c$=160 and 190\,MeV. In the former case, 
the total thermal yield for masses $M$=1-1.5\,GeV is indeed dominated
by QGP emission, while in the latter case hadronic emission prevails
(with chiral mixing, cf.~Fig.~\ref{fig_VA-mix}). The corresponding
two curves in a pair essentially overlap. A more significant effect 
is due to transverse flow. The original predictions of 
fireball~\cite{vanHees:2006ng,Ruppert:2007cr} and
hydrodynamic~\cite{Dusling:2006yv}
evolutions agree with each other, but underestimate the 
data~\cite{Damjanovic:2007qm} (cf.~the lower pair of curves in
Fig.~\ref{fig_slopes} right). When implementing a stronger transverse 
expansion~\cite{vanHees:2007th}, comparable to that used at RHIC 
($a_t$=0.1~fm/$c^2$), a reasonable description of the NA60 slopes can 
be obtained (upper two curves in Fig.~\ref{fig_slopes} right). Using 
the schematic slope formula with $T$=175\,MeV and an average QGP+mixed
phase evolution time of 3\,fm/$c$ for semicentral In-In (resulting
in $\langle\beta\rangle\simeq\frac{2}{3} a_t \tau = 0.2c$) gives an inverse
slope of $T_{\rm eff}\simeq 215$\,MeV for $M\simeq1$\,GeV, as seen
in the figure. The main point is that the NA60 slopes for $M\ge1$\,GeV
give an independent confirmation of thermal emission from a QCD
medium with dual rate and temperature of around $T_c$.     
The sensitivity to the unexpectedly large collective flow shows that
dileptons can also serve as an accurate ``barometer" of the fireball.   
A similar analysis at RHIC should lead to much increased slope 
parameters, since the QGP has built up much more flow by the time
it has cooled down to $T_c$. For example, from the left panel of
Fig.~\ref{fig_slopes} one finds $\beta_s\simeq0.6$fm/$c^2$ at the
end of the mixed phase, translating into $T_{\rm eff}\simeq350$\,MeV,
see also Refs.~\cite{Deng:2010pq,Ghosh:2010wt}.
Finally we remark that invariant-mass (rather than transverse-momentum)
dilepton spectra can, in principle, provide a cleaner measurement of the 
temperature since it is Doppler-free. Of course, the richer (dynamic) 
structure in the low-mass regime complicates this task, but with a 
reliable knowledge of the in-medium spectral shape it is still possible;
for the structureless regime above $M$=1\,GeV the situation is even 
better. Analyzing the invariant-mass slopes of the NA60 spectra
one finds temperatures ranging from 150-180\,MeV for 
$M\simeq0.2-1.5$\,GeV, approaching 200\,MeV for $M\ge2$\,GeV. 
The latter regime requires a good knowledge of the primordial
contribution, i.e., Drell-Yan annihilation and its nuclear 
corrections~\cite{Qiu:2001zj}.

\section{Conclusions}
\label{sec_concl}
Electromagnetic emission spectra provide a rich observable to 
investigate the strongly interacting medium produced in high-energy
collisions of heavy nuclei. The starting point for exploiting this
observable is a thorough understanding of the vector spectral function
in equilibrium matter. Chiral effective models at low and intermediate
temperature and densities, lattice-QCD in the vicinity of the phase 
transition(s) and perturbative QCD in the high-temperature limit
are starting to fill the theoretical landscape. The evidence is 
mounting for a melting of the hadronic vector-meson resonances and 
a transition to a partonic-like rate over the entire mass range.
This, in particular, includes a large low-mass enhancement which is 
corroborated by recent lattice QCD results for euclidean correlators 
and has been consistently showing in dilepton measurements from
BEVALAC/SIS via SPS to RHIC. The state-of-the art in dilepton 
experimentation has been set by the NA60 dimuon results, which 
accurately confirm the resonance melting and show that it occurs right 
around the temperatures where lQCD predicts the chiral phase transition.    
This sets the stage for exciting developments to come, at both collider 
energies (RHIC and LHC, where the connections to lQCD are even more 
direct) and at lower energies as planned at the Compressed Baryonic 
Matter experiment (CBM, where the baryon-induced medium effects are 
expected to become maximal). Several theoretical developments will have 
to accompany these efforts, including hadronic calculations of the
axialvector to tighten the grip on chiral restoration using sum rules, 
lattice calculations of euclidean correlators with dynamical quarks,
and the implementation of equilibrium rates into state-of-the-art 
hydrodynamic models (possibly supplemented with transport simulations) 
which quantitatively describe bulk-hadron observables.   
One can thus hope that the systematic use of EM probes will continue 
to deliver unique and precise information on the phase structure of a 
strongly coupled non-abelian gauge theory. 
\\ 

{\bf Acknowledgments}\\
I thank H.~van Hees, C.~Gale and J.~Wambach for longterm fruitful 
collaboration on the presented topics, and Z.~Fodor, L.~McLerran, 
R.~Pisarski and C.~Ratti for interesting discussion.
This work is supported by the US National Science Foundation under grant 
no.~PHY-0969394 and by the A.-v.-Humboldt foundation. 


\end{document}